\documentclass[preprint2,times]{aastex631}

\PassOptionsToPackage{dvipsnames,svgnames,x11names}{xcolor}
\usepackage{tabularx}

\usepackage{amsmath}  
\usepackage{amssymb}  
\usepackage[T1]{fontenc}
\usepackage{xspace}
\usepackage{multirow}
\usepackage{xcolor}
\usepackage{xspace}
\usepackage[mathlines]{lineno}
\usepackage[title]{appendix}
\usepackage{newunicodechar}
\newunicodechar{−}{-}

\defcitealias{kenworthy_salt3_2021}{K21}

\usepackage{newunicodechar,graphicx}
\DeclareRobustCommand{\okina}{%
  \raisebox{\dimexpr\fontcharht\font`A-\height}{%
    \scalebox{0.8}{`}%
  }%
}
\newunicodechar{ʻ}{\okina}

\newif{\ifchangetext}
\changetextfalse

\InputIfFileExists{\jobname-options}

\ifchangetext
  
  \newcommand{\changenote}[1]{\textcolor{blue}{ \bf #1}}x
\else
  
  \newcommand{\changenote}[1]{}
\fi

\hypersetup{
    colorlinks=True,
    citecolor=black,
    linkcolor=black,
    filecolor=magenta,      
    urlcolor=cyan,
}

\def\arcsec{\ensuremath{^{\prime\prime}}}

\def\xone{\ensuremath{x_{1}}}
\def\C{\ensuremath{\mathcal{C}}}

\newcommand{\lensedsn}{SN\,Encore\xspace}

\newcommand{\sntd}{{\fontfamily{qcr}\selectfont{SNTD}}\xspace}

\newcommand{\hst}{\textit{HST}\xspace}

\newcommand{\webb}{\textit{JWST}\xspace}

\newcommand{\STScI}{Space Telescope Science Institute, Baltimore, MD 21218, USA}
\newcommand{\JHU}{Physics and Astronomy Department, Johns Hopkins University, Baltimore, MD 21218, USA}
\newcommand{\UOArizona}{Department of Astronomy/Steward Observatory, University of Arizona, 933 N. Cherry Avenue, Tucson, AZ 85721, USA}
\newcommand{\arizonastate}{School of Earth and Space Exploration, Arizona State University, Tempe, AZ 85287-6004, USA}

\newcommand{\NEF}{NASA Einstein Fellow}
\newcommand{\milan}{Dipartimento di Fisica, Universit\`a degli Studi di Milano, via Celoria 16, 20133 Milano, Italy}
\newcommand{\ferrara}{Dipartimento di Fisica e Scienze della Terra, Universit\`a degli Studi di Ferrara, via Saragat 1, 44122 Ferrara, Italy} 
\newcommand{\portsmouth}{Institute of Cosmology and Gravitation, University of Portsmouth, Burnaby Rd, Portsmouth PO1 3FX, UK}
\newcommand{\negev}{Physics Department, Ben-Gurion University of the Negev, PO Box 653, Be'er-Sheva 84105, Israel}
\newcommand{\tum}{Technical University of Munich, TUM School of Natural Sciences, Physics Department,  James-Franck-Stra{\ss}e 1, 85748 Garching, Germany}
\newcommand{\mpi}{Max-Planck-Institut f{\"u}r Astrophysik, Karl-Schwarzschild Stra{\ss}e 1, 85748 Garching, Germany\\
}

\newcommand{\sntdstardtba}{$-39.8$}
\newcommand{\sntdstardtbaerr}{$^{+3.9}_{-3.3}$}

\usepackage{xspace}
\usepackage[mathlines]{lineno}
\begin{document}

\title{Cosmology with supernova Encore in the strong lensing cluster MACS J0138$−$2155: Time delays \& Hubble constant measurement}

\author[0000-0002-2361-7201
]{J.~D.~R.~Pierel}
\correspondingauthor{J.~D.~R.~Pierel} 
\email{jpierel@stsci.edu}
\altaffiliation{\NEF}
\affiliation{\STScI}

\author[0000-0003-3847-0780]{E.~E.~Hayes}
\affiliation{Institute of Astronomy and Kavli Institute for Cosmology, University of Cambridge, Madingley Road, Cambridge CB3 0HA, UK}

\author[0000-0001-7051-497X]{M.~Millon}
\affiliation{Institute for Particle Physics and Astrophysics, ETH Zurich, Wolfgang-Pauli-Strasse 27, CH-8093 Zurich, Switzerland}

\author[0000-0003-2037-4619]{C.~Larison}
\affiliation{\STScI}
\affiliation{Department of Physics and Astronomy, Rutgers University, 136 Frelinghuysen Road, Piscataway, NJ 08854, USA}

\author[0009-0009-7962-656X]{E.~Mamuzic}
\affiliation{\tum}
\affiliation{\mpi}

\author[0000-0003-3108-9039]{A.~Acebron}
\affiliation{Instituto de Física de Cantabria (CSIC-UC), Avda. Los Castros s/n, 39005 Santander, Spain}
\affiliation{INAF -- IASF Milano, via A. Corti 12, I-20133 Milano, Italy}

\author[0009-0008-1965-9012]{A.~Agrawal}
\affiliation{Department of Astronomy, University of Illinois Urbana-Champaign, 1002 West Green Street, Urbana, IL 61801, USA}
\affiliation{SkAI Institute, Chicago, IL, USA}

\author[0000-0003-1383-9414]{P.~Bergamini}
\affiliation{\milan}
\affiliation{INAF-Osservatorio di Astrofisica e Scienza dello Spazio di Bologna, Via Piero Gobetti 93/3, 40129 Bologna, Italy}

\author[0000-0001-7148-6915]{S.~Cha}
\affiliation{Department of Astronomy, Yonsei University, 50 Yonsei-ro, Seoul 03722, Korea}

\author[0000-0002-2376-6979]{S.~Dhawan}
\affiliation{School of Physics \& Astronomy and Institute of Gravitational Wave Astronomy, University of Birmingham, UK}

\author[0000-0001-9065-3926]{J.~M.~Diego}
\affiliation{Instituto de Fisica de Cantabria (CSIC-UC), Avda. Los Castros s/n, 39005, Santaner, Spain}

\author[0000-0003-1625-8009]{B.~L.~Frye}
\affiliation{\UOArizona}

\author[0000-0002-5116-7287]{D.~Gilman}
\affiliation{Department of Astronomy and Astrophysics, University of Chicago, Chicago, IL 60637, USA}

\author[0000-0002-9512-3788]{G.~Granata}
\affiliation{\milan}
\affiliation{\ferrara}
\affiliation{\portsmouth}

\author[0000-0002-5926-7143]{C.~Grillo}
\affiliation{\milan}
\affiliation{INAF -- IASF Milano, via A. Corti 12, I-20133 Milano, Italy}

\author[0000-0002-5751-3697]{M.~J.~Jee}
\affiliation{Department of Astronomy, Yonsei University, 50 Yonsei-ro, Seoul 03722, Korea}
\affiliation{Department of Physics and Astronomy, University of California, Davis, One Shields Avenue, Davis, CA 95616, USA}

\author[0000-0001-9394-6732]{P.~S.~Kamieneski}
\affiliation{\arizonastate}

\author[0000-0002-6610-2048]{A~M.~Koekemoer}
\affiliation{\STScI}

\author[0000-0002-7876-4321]{A.~K.~Meena}
\affiliation{\negev}
\affiliation{Department of Physics, Indian Institute of Science, Bangalore 560012, India}

\author[0000-0001-7769-8660]{A.~B.~Newman}
\affiliation{Observatories of the Carnegie Institution for Science, 813 Santa Barbara Street, Pasadena, CA 91101, USA}

\author[0000-0003-3484-399X]{M.~Oguri}
\affiliation{Center for Frontier Science, Chiba University, 1-33 Yayoicho, Inage, Chiba 263-8522, Japan}
\affiliation{Department of Physics, Graduate School of Science, Chiba University, 1-33 Yayoicho, Inage, Chiba 263-8522, Japan}

\author[0000-0003-0209-9246]{E.~Padilla-Gonzalez}
\affiliation{\JHU}

\author[0000-0002-5391-5568]{F.~Poidevin}
\affiliation{Instituto de Astrof\'{\i}sica de Canarias, V\'{\i}a   L\'actea, 38205 La Laguna, Tenerife, Spain}
\affiliation{Universidad de La Laguna, Departamento de Astrof\'{\i}sica,  38206 La Laguna, Tenerife, Spain}

\author[0000-0002-6813-0632]{P.~Rosati}
\affiliation{\ferrara}

\author[0000-0003-2497-6334]{S.~Schuldt}
\affiliation{\milan}
\affiliation{INAF - IASF Milano, via A. Corti 12, I-20133 Milano, Italy}

\author[0000-0002-7756-4440]{L.~G.~Strolger}
\affiliation{\STScI}

\author[0000-0001-5568-6052]{S.~H.~Suyu}
\affiliation{\tum}
\affiliation{\mpi}

\author[0009-0005-6323-0457]{S.~Thorp}
\affiliation{Institute of Astronomy and Kavli Institute for Cosmology, University of Cambridge, Madingley Road, Cambridge CB3 0HA, UK}
\affiliation{The Oskar Klein Centre, Department of Physics, Stockholm University, AlbaNova University Centre, SE 106 91 Stockholm, Sweden}

\author[0000-0002-0350-4488]{A.~Zitrin}
\affiliation{\negev}

\vspace{-10pt}

\begin{abstract}

\noindent Multiply-imaged supernovae (SNe) provide a novel means of constraining the Hubble constant ($H_0$). Such measurements require a combination of precise models of the lensing mass distribution and an accurate estimate of the relative time delays between arrival of the multiple images. Only two multiply-imaged SNe, Refsdal and H0pe, have enabled measurements of $H_0$ thus far. Here we detail the third such measurement for \lensedsn, a $z=1.95$ SN\,Ia discovered in {\it JWST}/NIRCam imaging. We measure the time delay, perform simulations of additional microlensing and millilensing systematics, and combine with the mass models of Suyu et al. in a double-blind analysis to obtain our $H_0$ constraint.  Our final time-delay measurement is $\Delta t_{1b,1a}=-39.8_{-3.3}^{+3.9}$ days, which is combined with seven lens models weighted by the likelihood of the observed multiple image positions for a result of $H_0=66.9_{-8.1}^{+11.2}\,\rm{km}\,\rm{s}^{-1}\rm{Mpc}^{-1}$. The uncertainty on this measurement could be improved significantly if template imaging is obtained. Remarkably, a sibling to \lensedsn (SN\,“Requiem”) was discovered in the same host galaxy, making the MACS J0138.0−2155 cluster the first system known to produce more than one observed multiply-imaged SN. SN\,Requiem has a fourth image that is expected to appear within a few years, providing an unprecedented decade-long baseline for time-delay cosmography and an opportunity for a high-precision joint estimate of $H_0$. 

\end{abstract}

\section{Introduction}
\label{sec:intro}
The Hubble Constant ($H_0$) is a critical parameter for cosmological models, as it sets the size and age scale of the universe. The most precise local measurement of $H_0$ leverages luminosity distances measured from Type Ia supernovae (SNe\,Ia), in conjunction with a local distance ladder approach, resulting in $H_0=73.04\pm1.04\,\rm{km}\,\rm{s}^{-1}\,\rm{Mpc}^{-1}$ \citep{riess_comprehensive_2022}. Conversely the value of $H_0$ has been predicted with a cosmological model (e.g., the static dark energy and cold dark matter, $\Lambda$CDM) calibrated with measurements of the cosmic microwave background (CMB) in the early universe, resulting in $H_0=67.4\pm0.5\,\rm{km}\,\rm{s}^{-1}\,\rm{Mpc}^{-1}$ \citep{planck_collaboration_planck_2020}. This significant discrepancy, known as the Hubble Tension, is a major topic of research in cosmology today. Either missing systematics from the local measurement, or physics missing from the $\Lambda$CDM model, would have significant implications for our understanding of the universe and future cosmological measurements.

One avenue to resolving the Hubble Tension is to perform fully independent measurements of $H_0$, thereby providing an additional constraint on its true value. This has been an active effort in recent years, with a wide range of alternative probes (See section \ref{sec:results} for a figure) including early universe baryon acoustic oscillation (BAO) and Big Bang nucleosynthesis (BBN) \citep[][$H_0=67.6\pm1.0$]{schoneberg_baobbn_2022}, dark sirens \citep[][$H_0=72.6^{+4.1}_{-4.2}$]{wang_late_2023}, megamasers \citep[][$H_0=73.9\pm3.0$]{pesce_megamaser_2020}, and other distance ladder approaches such as Miras \citep[][$H_0=73.3\pm4.0$]{huang_hubble_2020}, surface brightness fluctuations (SBF) with SNe \citep[][$H_0=74.6\pm2.9$]{garnavich_connecting_2023} or tip of the red giant branch (TRGB) $+$ Cepheids \citep[][$H_0=73.3\pm2.5$, $70.4\pm1.9$]{blakeslee_hubble_2021,freedman_status_2025}. These methods essentially span the width of the Hubble Tension and many have correlated or shared systematics (see Section \ref{sec:results} for a graphical representation). Additional methods that are unique to these and can reach high-precision are therefore desirable. 

Precise measurements of the ``time delay'' between the arrival of a multiply-imaged source yield a direct distance measurement to the lens system that constrains the Hubble constant ($H_0$) in a single step \citep[e.g.,][]{refsdal_possibility_1964,linder_lensing_2011,paraficz_gravitational_2009,treu_time_2016,grillo_measuring_2018,grillo_accuracy_2020,birrer_time-delay_2022,treu_strong_2022,kelly_constraints_2023,grillo_cosmography_2024,suyu_strong_2024}. Strong gravitational lensing can cause these multiple images of a background source to appear, as light propagating along different paths are focused by the gravity of a foreground galaxy or galaxy cluster \citep[called the ``lens''; e.g.,][]{narayan_lectures_1997}. Such a phenomenon requires chance alignment between the observer, the background source, and the lens. If the multiply-imaged source has variable brightness then depending on the relative geometrical and gravitational potential differences of each path, the source images will be measurably delayed by hours to weeks (for galaxy-scale lenses, $M\lesssim10^{12}M_\odot$) or months to years \citep[for cluster-scale lenses, $M\gtrsim10^{13}M_\odot$; e.g.,][]{oguri_strong_2019}. While such ``time-delay cosmography'' has been accomplished with quasars \citep[e.g.,][the latest measurement by the TDCOSMO collaboration yield $H_0 = 71.6^{+3.9}_{−3.3} \, \rm{km}\,\rm{s}^{-1}\, \rm{Mpc}^{-1}$, using a sample of 8 lensed quasars]{kundic_robust_1997,schechter_quadruple_1997,burud_time_2002,hjorth_time_2002,vuissoz_cosmograil_2008,suyu_dissecting_2010,tewes_cosmograil_2013,bonvin_h0licow_2017,bonvin_cosmograil_2018,bonvin_cosmograil_2019,birrer_h0licow_2019,chen_sharp_2019,wong_h0licow_2020,birrer_tdcosmo_2020,millon_tdcosmo_2020,shajib_strides_2020,shajib_tdcosmo_2023,tdcosmo_collaboration_tdcosmo_2025}, multiply-imaged SNe have been slower to materialize despite a variety of valuable characteristics, and advantages over quasars including predictable evolution, relative simplicity of observations, standardizable brightness (for SNe\,Ia), and mitigated microlensing effects \citep[e.g.,][]{foxley-marrable_impact_2018,goldstein_precise_2018,pierel_turning_2019,huber_holismokes_2021,ding_improved_2021,pierel_projected_2021,birrer_hubble_2022,chen_jwst_2024,pierel_jwst_2024,pascale_sn_2025}. 

The sample of strongly lensed SNe has grown over the last decade to include two SNe\,Ia in galaxy-scale systems \citep{goobar_iptf16geu:_2017,goobar_uncovering_2023,pierel_lenswatch_2023} and six cluster-scale systems \citep{kelly_multiple_2015,rodney_gravitationally_2021,chen_jwst-ers_2022,kelly_strongly_2022,frye_jwst_2024,pierel_lensed_2024}, though only SN Refsdal \citep{kelly_constraints_2023,kelly_magnificent_2023} and SN H0pe \citep{chen_jwst_2024,frye_jwst_2024,pascale_sn_2025,pierel_jwst_2024} have had sufficiently long time-delays and well-sampled light curves for $H_0$ constraints with relatively small uncertainty. SN Refsdal yielded a $\sim$6\% measurement of $H_0$ in flat $\Lambda$CDM cosmology \citep[with $\Omega_m=0.3$: 64.8$_{-4.3}^{+4.4}$ or 66.6$_{-3.3}^{+4.1 }$ km s$^{-1}$ Mpc$^{-1}$, depending on lens model weights;][]{kelly_constraints_2023}, 
and also $65.1_{-3.4}^{+3.5 }$ km s$^{-1}$ Mpc$^{-1}$ in a more general background cosmological model \citep{grillo_cosmography_2024}. SN H0pe is of particular interest as the first lensed SN\,Ia to provide an $H_0$ result competitive with local measurements, resulting in $75.4^{+8.1}_{-5.5}$ km s$^{-1}$ Mpc$^{-1}$ \citep{pascale_sn_2025}. 

\begin{figure*}[!ht]
    \centering
        \includegraphics[trim={0cm 4cm 0cm 0cm},clip,width=.99\linewidth]{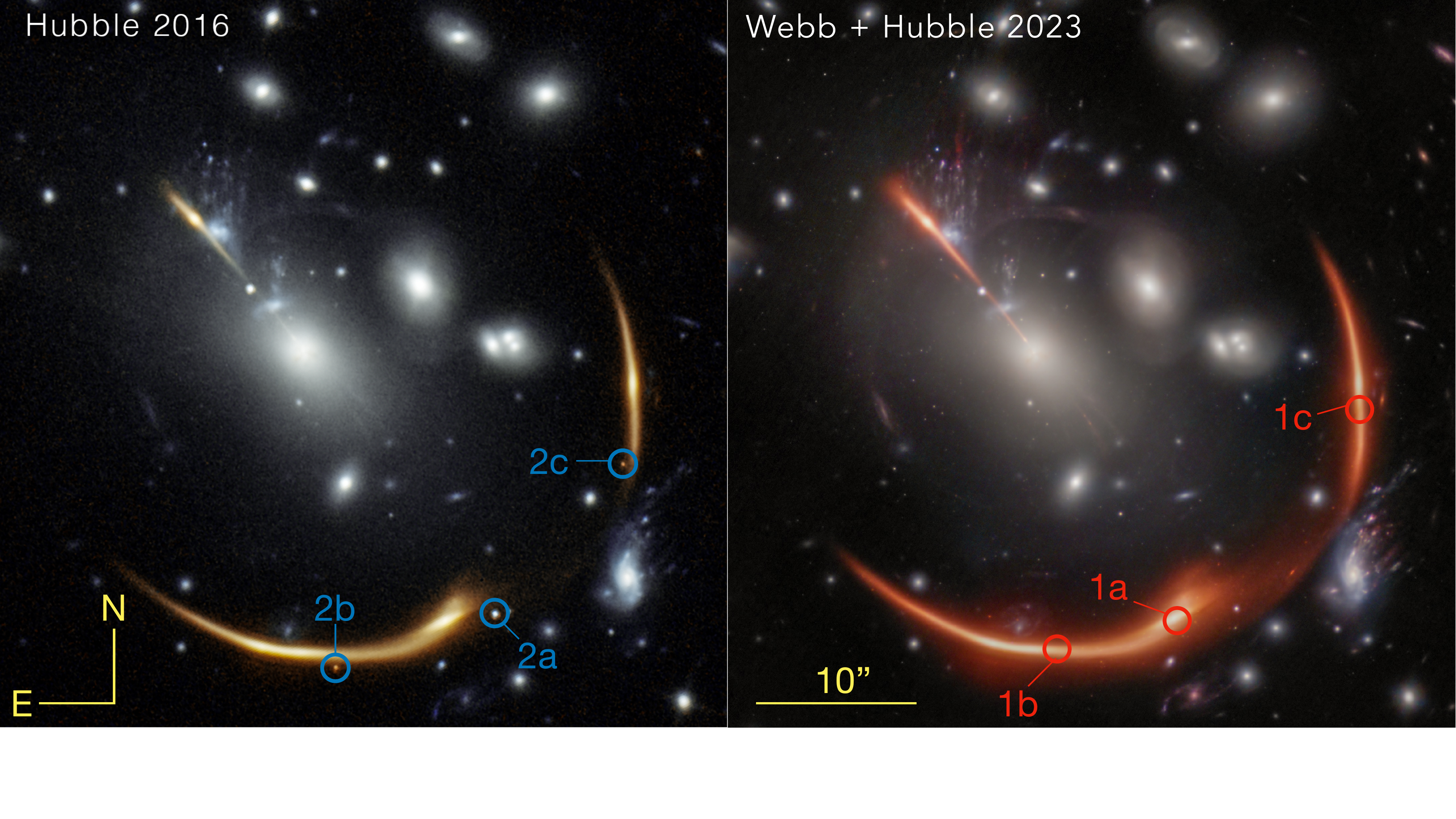}
    \caption{  {\it Left}: {\it HST} WFC3/IR two-color image in the region of MACS J0138.0$-$2155 from $2016$, using F105W (blue) and F160W (orange). SN Requiem is marked in its three visible image positions by blue circles, notably absent in $2023$. {\it Right:} Combined \textit{JWST}/NIRCam and \textit{HST}/WFC3 color image from programs 6549 and 16264 (Table \ref{tab:obs}). The filters used are F105W+F115W+F125W (blue), F150W+F160W+F200W (green), and F277W+F356W+F444W (red). The images were drizzled to $0\farcs02/\rm{pix}$, and the image scale and orientation are as shown. The three detected image positions of \lensedsn are circled in red, but it is not visible at this scale (See Figure \ref{fig:starred_model} for a zoom in of positions 1a and 1b)  (Image Credit: STScI, A. Koekemoer, T. Li).}
    \label{fig:color_im}
\end{figure*}

SNe\,Ia are of particular value when strongly lensed, as their standardizable absolute brightness can provide additional leverage for lens modeling by limiting the uncertainty caused by the mass-sheet degeneracy \citep{falco_model_1985,kolatt_gravitational_1998,holz_seeing_2001,oguri_gravitational_2003,patel_three_2014,nordin_lensed_2014,rodney_illuminating_2015,xu_lens_2016,foxley-marrable_impact_2018,birrer_hubble_2022}, though only in cases where millilensing and microlensing are not extreme \citep[see][]{goobar_iptf16geu:_2017,yahalomi_quadruply_2017,foxley-marrable_impact_2018,dhawan_magnification_2019}. Additionally, SNe\,Ia have well-understood models of light curve evolution \citep{hsiao_k_2007,guy_supernova_2010,saunders_snemo_2018,leget_sugar_2020,kenworthy_salt3_2021,mandel_hierarchical_2022,pierel_salt3nir_2022} that enable precise time-delay measurements using color curves, which removes the effects of macro- and achromatic microlensing \citep[e.g.,][]{pierel_turning_2019,huber_holismokes_2021,rodney_gravitationally_2021,pierel_lenswatch_2023, grupa_holismokes_2025}. ``SN Requiem'' \citep{rodney_gravitationally_2021} was the first cluster-scale \textit{photometrically} classified multiply-imaged SN\,Ia, appearing at  $z=1.95$ in the MRG-M0138 galaxy that is lensed by the MACS J0138.0$-$2155 cluster (Figure \ref{fig:color_im}). Unfortunately, the SN was found archivally several years after it had faded, precluding an $H_0$ measurement with the visible images, though a fourth image of SN Requiem is expected to arrive in $\sim2026$-$2027$ after a decade delay (Suyu et al., submitted; hereafter S25). 
Remarkably, a second lensed SN\,Ia (dubbed \lensedsn) was later found in \textit{James Webb Space Telescope} (\textit{JWST}) imaging of the same SN Requiem host galaxy at $z=1.95$. The initial discovery is reported in \citet[][hereafter P24]{pierel_lensed_2024}, with the spectroscopic classification and analysis described in \citet{dhawan_spectroscopic_2024}. In addition to being the first discovery of a lensed SN sibling pair, SN Encore is notable as its estimated time delay and comprehensive follow-up campaign make it just the third lensed SN (second lensed SN\,Ia) capable of producing a competitive $H_0$ measurement.

This work measures $H_0$ with SN Encore, by analyzing the light curves to determine the relative time delay between images and combining with a series of mass model predictions. The observations used to make this measurement are described briefly in Section \ref{sec:obs}, followed by the methods used to extract photometry for each SN image in Section \ref{sec:phot}. The light curve fitting and time-delay measurement are described in Section \ref{sec:fitting}, and additional systematic uncertainties for the final results are explored in Section \ref{sec:sys}. The measurements are combined with mass models for the system from Suyu et al. (submitted) and final results are given in Section \ref{sec:results}, with a discussion on the significance of the results, upcoming improvements, and future prospects in Section \ref{sec:conclusion}.


\section{Summary of Observations}
\label{sec:obs}

The discovery of \lensedsn and subsequent follow-up is described in detail by P24. Briefly, \lensedsn was discovered by the \textit{JWST}-GO-2345 team in F150W NIRCam imaging taken 2023 November 17 (MJD $60265$) by comparing the data with an archival \hst WFC3/IR F160W image \citep[2016 July 18, MJD $57587$][]{newman_resolving_2018}. \citet{newman_resolving_2018} reported a spectroscopic redshift for MRG-M0138 of $z=1.95$, and the host properties alone suggested that \lensedsn was likely an SN\,Ia at $z=1.95$ \citep{rodney_gravitationally_2021}. Subsequently, follow-up observations using the \textit{Hubble Space Telescope} (\textit{HST}) and \textit{JWST} were triggered, as the only telescope resources with the required resolution and wavelength range to detect \lensedsn. Color images of the system, including both \hst and \textit{JWST} and our image labeling convention for both SN\,Requiem and \lensedsn, are shown in Figure \ref{fig:color_im}.  These observations are briefly described below, with details given in P24. 

\begin{table*}[!t]
    \centering
    \caption{\label{tab:obs} Summary of observations taken of SN Encore, taken from P24.}
    
    \begin{tabular*}{\linewidth}{@{\extracolsep{\stretch{1}}}*{7}{c}}
\toprule
Program ID&Obs. Type&Telescope&Instrument&MJD&\multicolumn{1}{c}{Filter/Grating}&\multicolumn{1}{c}{Exp. Time (s)}\\
\hline
2345&Imaging&\textit{JWST}&NIRCam&$60265$&F150W&$773$\\
2345&Imaging&\textit{JWST}&NIRCam&$60265$&F444W&$773$\\
2345&Spectroscopy&\textit{JWST}&NIRSpec IFU&$60266$&G235M&$7586$\\
2345&Spectroscopy&\textit{JWST}&NIRSpec IFU&$60305$&G140M&$8170$\\

\hline
6549&Imaging&\textit{JWST}&NIRCam&$60283$&F115W&$1417$\\
6549&Imaging&\textit{JWST}&NIRCam&$60283$&F150W&$859$\\
6549&Imaging&\textit{JWST}&NIRCam&$60283$&F200W&$859$\\
6549&Imaging&\textit{JWST}&NIRCam&$60283$&F277W&$859$\\
6549&Imaging&\textit{JWST}&NIRCam&$60283$&F356W&$859$\\
6549&Imaging&\textit{JWST}&NIRCam&$60283$&F444W&$1417$\\
6549&Imaging&\textit{JWST}&NIRCam&$60301$&F115W&$1589$\\
6549&Imaging&\textit{JWST}&NIRCam&$60301$&F150W&$1074$\\
6549&Imaging&\textit{JWST}&NIRCam&$60301$&F200W&$1074$\\
6549&Imaging&\textit{JWST}&NIRCam&$60301$&F277W&$1074$\\
6549&Imaging&\textit{JWST}&NIRCam&$60301$&F356W&$1074$\\
6549&Imaging&\textit{JWST}&NIRCam&$60301$&F444W&$1589$\\
6549&Imaging&\textit{JWST}&NIRCam&$60317$&F115W&$1589$\\
6549&Imaging&\textit{JWST}&NIRCam&$60317$&F150W&$1074$\\
6549&Imaging&\textit{JWST}&NIRCam&$60317$&F200W&$1074$\\
6549&Imaging&\textit{JWST}&NIRCam&$60317$&F277W&$1074$\\
6549&Imaging&\textit{JWST}&NIRCam&$60317$&F356W&$1074$\\
6549&Imaging&\textit{JWST}&NIRCam&$60317$&F444W&$1589$\\
\hline
16264&Imaging&\textit{HST}&WFC3/IR&$60294$&F105W&$6913$\\
16264&Imaging&\textit{HST}&WFC3/IR&$60294$&F125W&$4609$\\
16264&Imaging&\textit{HST}&WFC3/IR&$60294$&F160W&$2304$\\
16264&Imaging&\textit{HST}&WFC3/IR&$60339$&F105W&$9318$\\
16264&Imaging&\textit{HST}&WFC3/IR&$60339$&F125W&$4609$\\
16264&Imaging&\textit{HST}&WFC3/IR&$60339$&F160W&$2304$\\
\hline

\hline
    \end{tabular*}
\begin{flushleft}
\tablecomments{Columns are: \textit{JWST}/\textit{HST} Program ID, observation type, telescope name, instrument name, Modified Julian date, filter/grating, and exposure time in seconds.}

\end{flushleft}
\end{table*}

\subsection{\textit{HST} Data}
\label{sub:obs_hst}
LensWatch\footnote{\url{https://www.lenswatch.org}} is a collaboration with the goal of finding
gravitationally lensed SNe, both by monitoring active transient surveys \citep[e.g.,][]{fremling_zwicky_2020,jones_young_2021} and by
way of targeted surveys \citep{craig_targeted_2021}. The collaboration
maintained a Cycle 28 \textit{HST} program (HST-GO-16264) given long-term (three-cycle) target-of-opportunity (ToO) status. The program included three ToO
triggers (two non-disruptive, one disruptive), and was designed
to provide high-resolution follow-up imaging for a ground-based lensed SN discovery, which is critical for galaxy-scale
multiply imaged SNe due to their small image separations \citep[e.g.,][]{goobar_iptf16geu:_2017,goobar_uncovering_2023,pierel_lenswatch_2023}. LensWatch triggered \textit{HST}-GO program 16264 on 2023 November 20 (MJD $60268$, three days after discovery) to obtain follow-up imaging of \lensedsn, including $14$ orbits in three filters spread over two epochs (Table \ref{tab:obs}). The trigger was non-disruptive due to a technical disruption to \textit{HST} observations; the first epoch was scheduled $26$ ($\sim9$ rest-frame) days later on 2023 December 16 (MJD $60294$) and the second epoch $45$ days after the first on 2024 January 30 (MJD $60339$). These observations are separated by $10$-$22$ observer-frame ($3$-$7$ rest-frame) days from the \textit{JWST} observations described in the following section, and were designed to be complementary to \textit{JWST}. Details of the \textit{HST} data processing and photometry are given in Section \ref{sec:phot}.

All the HST data were retrieved from the STScI MAST archive, and the calibrated exposures were subsequently processed with additional improvements described here as follows. The default astrometry from the archive was corrected by realigning all the HST exposures to one another, as well as to the JWST images and to Gaia-DR3, using an updated version of the HST ``mosaicdrizzle'' pipeline first described in \citet{koekemoer_candels_2011}, which also removes residual low-level background variations across the detectors. This pipeline was then used to combine all the HST data into mosaics for each filter with all distortion removed, producing mosaics for the separate epochs at a pixel scale of $40$mas.

\subsection{\textit{JWST} Data}
\label{sub:obs_jwst}
\subsubsection{Imaging}
\label{sub:obs_jwst_img}
Despite the \textit{HST} observations outlined in the previous section, \textit{JWST} was still required to fully leverage \lensedsn for two reasons: 1) the small host separation ($\sim0.1\arcsec$) meant that even with difference imaging, it would be very difficult to robustly detect \lensedsn, and 2) based on the discovery epoch \lensedsn (image 1a) was between peak brightness and the second infrared (IR) maximum for SNe\,Ia \citep{hsiao_k_2007,krisciunas_carnegie_2017,pierel_salt3nir_2022}. At $z=1.95$, \textit{HST} is only able to cover rest-frame uBV filters, meaning \textit{JWST} was needed to reach the rest-frame near-IR and leverage the second infrared maximum for more accurate time-delay measurements.

A disruptive \webb director's discretionary time (DDT) proposal was subsequently approved (\textit{JWST}-GO-6549, PI Pierel), which provided three additional epochs of NIRCam imaging in six filters (Table \ref{tab:obs}). The first NIRCam epoch occurred on 2023 December 5 (MJD $60283$), $18$ ($\sim6$ rest-frame) days after discovery. The next two epochs took place with a cadence of $\sim17$ observer-frame ($\sim6$ rest-frame) days, on 2023 December 23 (MJD $60301$) and 2024 January 8 (MJD $60317$). This meant the \textit{JWST} light curve for each image contains four observations, each separated by $\sim6$ rest-frame days. The  summary of observations is given in Table \ref{tab:obs}, and \textit{JWST} data processing and photometry are described in Section \ref{sec:phot}.  

All the NIRCam data were retrieved from the STScI MAST\footnote{\url{https://mast.stsci.edu}} archive, and were processed and calibrated using the JWST Pipeline\footnote{\url{https://github.com/spacetelescope/jwst}} 
\citep{bushouse_jwst_2022} version 1.12.5, with CRDS reference files defined in 1180.pmap, with additional improvements described here as follows. As part of the initial processing with the Stage 1 portion of the pipeline, additional corrections were applied to remove 1/f noise as well as low-level background variations across the detectors, including the removal of wisps and other low-level artifacts, with these techniques all described in more detail in \citet{windhorst_jwst_2023}. The Stage 2 portion of the pipeline was then run to apply photometric calibration to physical units, using the latest photometric calibrations \citep{boyer_jwst_2022}.

The images were then all astrometrically aligned directly to the Gaia-DR3\footnote{\url{https://www.cosmos.esa.int/web/gaia/dr3}} reference frame and finally combined into mosaics for each filter with all distortion removed, using the Stage 3 portion of the pipeline, where mosaics were constructed for each of the individual epochs. These mosaics have been drizzled to a $40$mas pixel scale for all filters, and are on the same pixel grid as the \textit{HST} mosaics from the previous section.

\subsubsection{Spectroscopy}
\label{sub:obs_jwst_spec}
\begin{figure*}
    \centering
    \includegraphics[trim={2.5cm 0.5cm 3.5cm 0.5cm},clip,width=\linewidth]{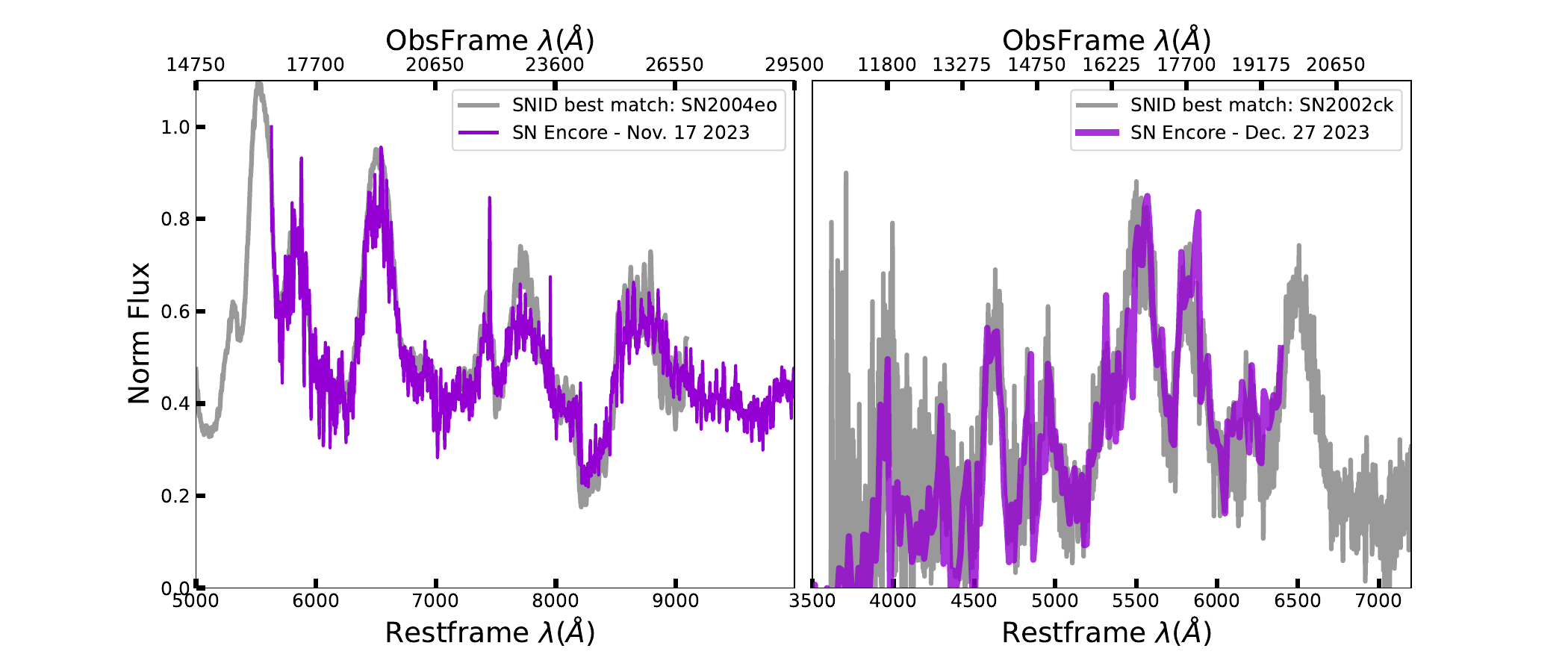}
    \caption{The observed G235M (left) and G140M (right) NIRSpec spectra of SN Encore are shown in purple, with the best-fit \texttt{SNID} templates shown in grey. Figure adapted from \citet{dhawan_spectroscopic_2024}.}
    \label{fig:encore-spec}
\end{figure*}

The \textit{JWST}-GO-2345 program also included NIRSpec integral field unit (IFU) spectroscopy for image 1a of MRG-M0138 using both the G140M and G235M gratings. There was a failure in the G140M observation causing it to be scheduled $39$ days later on 2023 December 27, giving some temporal information from the spectroscopy. \citet{dhawan_spectroscopic_2024} performed a detailed extraction and analysis of these spectra, concluding that \lensedsn was a somewhat fast-declining SN\,Ia $\sim30$ rest-frame days after peak brightness at the time of discovery. The spectra of \lensedsn, and their comparison to previous normal SNe\,Ia, are shown in Figure \ref{fig:encore-spec}. We leverage the fact that \lensedsn is spectroscopically confirmed as a SN\,Ia, as well as the spectroscopic phase inference, in Section \ref{sec:fitting}.

\section{Measuring Photometry of SN Encore }
\label{sec:phot}
No template images yet exist for the \textit{JWST} NIRCam imaging of SN Encore, making direct subtraction of the host galaxy light impossible, and so we must rely on alternative methods. We measure two different sets of photometry, as each method has some advantages and disadvantages, then measure time delays on each set in Section \ref{sec:fitting} for comparison. The goal of each method is to measure (at least) two of the visible images of SN Encore in as many NIRCam filters as possible, in order to enable robust time-delay measurements. These methods are described below. Both photometry methods make use of Level $3$ mosaic products described in the previous section and P24. Measured photometry tables, as well as some ancillary information, are given in the appendix.

\subsection{Direct Background Estimation}
\label{sub:phot_bkg}
\begin{figure}
    \centering
    \includegraphics[width=.8\linewidth,trim={4cm 0cm 2cm 0cm},clip]{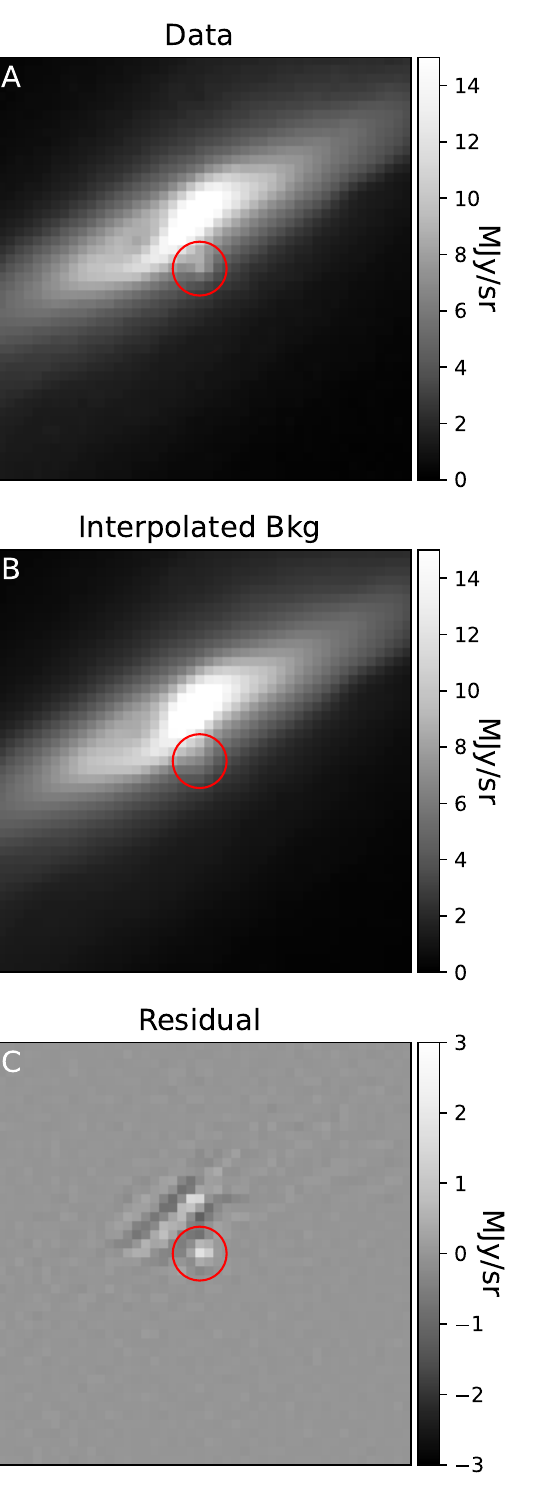}
    \caption{An example of the direct background estimation method in the F150W filter (epoch 1). The observed data is on the top, followed by the estimated background from the 2D polynomial method (middle), and resulting residual (bottom). }
    \label{fig:bkg_interp}
\end{figure}

We first attempt to measure the flux of SN Encore at each image position using the \texttt{AstroBkgInterp} package\footnote{https://github.com/brynickson/AstroBkgInterp}, which was developed specifically for estimating and removing complex backgrounds around point sources in \textit{JWST} data. The package masks the known position of a point source in the image, then fits a 2D polynomial to the surrounding region of the image and subtracts that result from the original image, leaving only the masked point source (plus residuals). We use the 2D polynomial method and a masking box of $5\times5$ pixels, with an overall region that is $49\times49$ pixels ($\sim1.5\arcsec$ on a side) centered on the SN position (Figure \ref{fig:bkg_interp}). We run this method on all filters and epochs at all three image positions of SN Encore (1a, 1b and 1c, see Figure \ref{fig:color_im}), then measure point-spread function (PSF) photometry. For PSF models we follow the methods used for SN H0pe \citep{pierel_jwst_2024}, which begins by using STPSF (version 2.0.0)\footnote{\url{https://www.stsci.edu/jwst/science-planning/proposal-planning-toolbox/psf-simulation-tool}} to represent the (level 2) PSF, which has been updated to better match the observed PSF in each filter, and takes into account temporal and spatial variation across the detector. We then drizzle these PSF models at the position of the SN using the same method that produced the drizzled data products, which produces an approximate PSF model usable on drizzled data. This process is repeated for all filters and epochs, and separately for each SN image position. 

Using the background-subtracted images produced above (Figure \ref{fig:bkg_interp}c) and the \texttt{space\_phot} package \citep{pierel_space-phot_2024}, we measure PSF photometry for all filters, epochs, and image positions. We then implement a PSF planting and recovery routine, where for all measurements we plant six PSF models at other positions around the host galaxy with similar flux levels, then recover the fluxes using the same routine described above. We find that this method is only reliable with the short-wavelength NIRCam filters (F115W, F150W, F200W) because of the brightness of the host galaxy and decreased resolution at longer wavelengths, and in these filters we recover a systematic uncertainty of $\sim0.1$ mag for images 1a and 1b. Image C is not robustly recovered in any filter with this method. We add this uncertainty in quadrature with the statistical uncertainty measured for each photometric data point, and proceed with only these three filters and images 1a and 1b for this method. The photometry measured in this section we will refer to as \texttt{phot\_bkg}.

Note that for reasons similar to the failure of this method for \textit{JWST} long-wavelength ($>2\mu$m) images, we do not attempt to measure \textit{HST} photometry in this section. 

\subsection{STARRED}
\label{sub:phot_starred}
STARRED \citep{Michalewicz2023, Millon_2024} is a Python package for PSF photometry optimized for light-curve extraction, implemented in JAX \citep{jax}. It uses a forward-modeling approach to isolate the point source components, which can be time-variable, from the extended background, which is assumed to be constant over time. This approach is often referred to as \textit{scene modeling}. In STARRED, the host galaxy is modeled on a grid of pixels with half the pixel size of the original image and regularized using starlets \citep{starck_starlet_2015}.

STARRED takes as input all time-series images along with a PSF model for each epoch and returns as output: i) the accurate astrometry for the point sources, ii) the photometry of the point sources for each input image, iii) translation shifts between images (due to telescope dithering and/or alignment errors) and, iv) a high-resolution model of the extended "background" sources. STARRED allows the user to choose the \textit{target PSF} of the high-resolution model, which is set to be a Gaussian with a full width at half maximum (FWHM) of two pixels. This choice is arbitrary but convenient, as it satisfies the Nyquist–Shannon sampling theorem, avoiding the formation of artifacts when reconstructing the high-resolution image \citep[see Fig. 1 of][for an illustration]{Millon_2024}.

The PSF is reconstructed with STARRED for each epoch using 3 or 4 non-saturated stars in the NIRCam field of view, depending on the filter. Our PSF model for the 4$^{\text{th}}$ epoch in six different JWST filters is shown in the appendix.

Once a PSF model is obtained for each epoch, we run STARRED on the F150W band first to obtain the most accurate estimate of the astrometric position of the SN images (1a and 1b), as this band has the highest signal-to-noise ratio. We then use the measured position as a Gaussian prior -- with a width of 0.05 pixel -- for the point source position and re-run the algorithm on all the other bands. This prior is especially important for the long-wavelength ($>2\mu$m) bands, where the SN is barely detected due to the glare of the much brighter host galaxy.

\begin{figure*}
    \centering
    \includegraphics[width=0.8\linewidth]{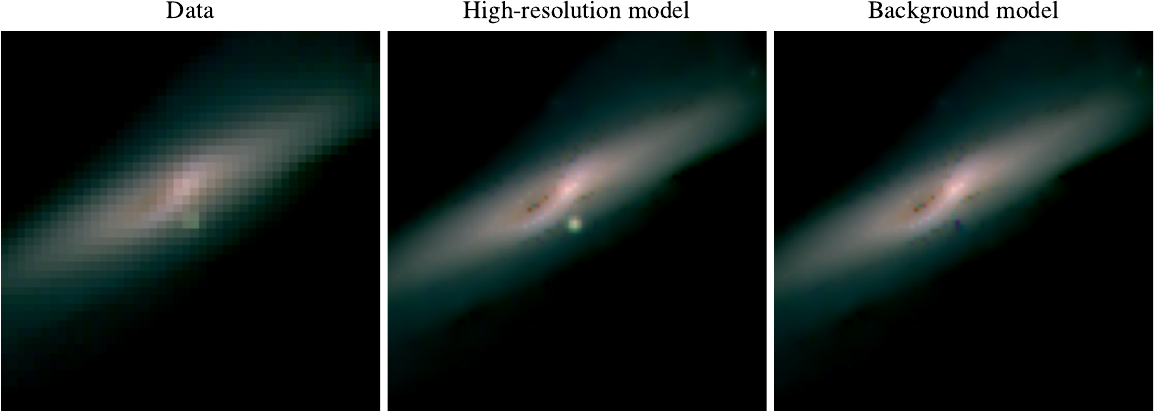}
    \includegraphics[width=0.8\linewidth]{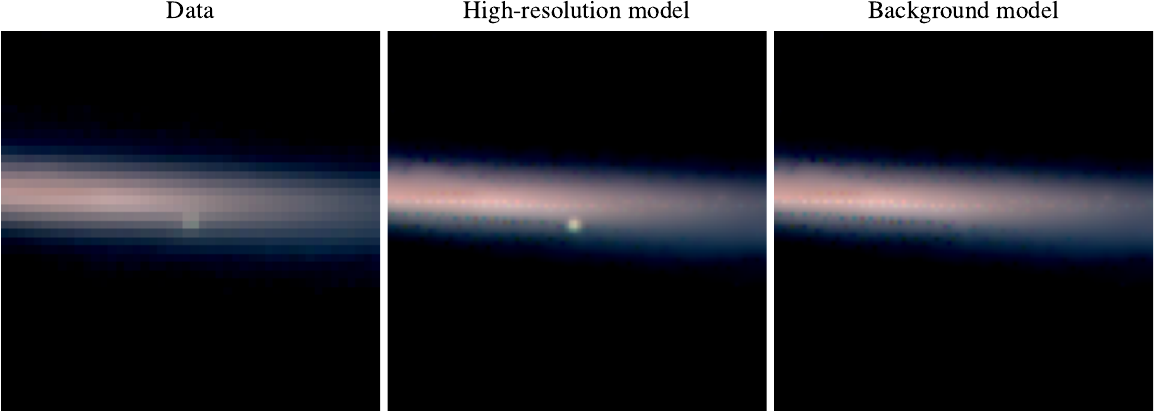}
    \caption{Left: Color image (R: F200W, G: F150W, B: F115W) constructed from the first epoch of our JWST observations for image 1a (top) and image 1b (bottom).
    Middle: High-resolution model produced by STARRED. The pixel size is 0.02"/pixel, subsampled by a factor of two relative to the original data. The PSF of this image is a Gaussian with a full width at half maximum (FWHM) of 2 pixels in the subsampled grid.
    Right: Background model with the SN flux—modeled as a point source—subtracted.
    \label{fig:starred_model}
    }
    
\end{figure*}

All bands are processed independently, as the host galaxy brightness is expected to vary significantly across the different JWST photometric bands. A color image constructed from the F115W, F150W, and F200W original data (first epoch) and the corresponding STARRED model (combining all epochs) are shown in Figure~\ref{fig:starred_model}. 

The photometry is extracted from the amplitudes of the point source for each epochs and for each band. For the \textit{HST} data, where a template image is available, we fix the amplitude of the point source to zero at the first observing epoch—taken before the SN explosion—and marginalize over several choices of the starlet regularization strength \citep[$\lambda = 1$–$5$, see][for details]{Millon_2024} to account for residual systematic deblending errors due to this choice of model hyperparameter.

For the \textit{JWST} data, where a template image is not yet available, we fix $\lambda = 3$ and account for possible systematic deblending errors arising from the degeneracy between the point source and the extended flux from the host galaxy when modeling the light curves (see Section \ref{sub:fitting_sntd}). The photometry measured in this section we will refer to as \texttt{phot\_starred}.

\section{Measurement of the Time Delay}
\label{sec:fitting}

In order to guard against potential biases, we measure the \lensedsn time delays using multiple methods and both sets of photometry (\texttt{phot\_bkg} and \texttt{phot\_starred}) from the previous section. We list these methods in order of least to most precise, with each step providing an additional robustness and systematics check. 

\subsection{SN\,Ia Light Curve Model}
\label{sub:fitting_models}

While the most precise time-delay measurements come from lensed SNe observed before peak brightness \citep{pierel_turning_2019}, a lensed SN\,Ia observed after peak brightness can still be robustly measured \citep{pierel_turning_2019} using well-calibrated SED models of SN\,Ia light curve evolution and variability \citep[e.g.,][]{guy_salt2:_2007,hsiao_k_2007,pierel_extending_2018,leget_sugar_2020,kenworthy_salt3_2021,mandel_hierarchical_2022,pierel_salt3nir_2022}. Of these, the most well-studied is the Spectral Adaptive Lightcurve Template (SALT) model \citep{guy_salt:_2005,guy_salt2:_2007,guy_supernova_2010,betoule_improved_2014,pierel_extending_2018,kenworthy_salt3_2021,pierel_salt3nir_2022}, which extends to $50$ rest-frame days after peak brightness. As described in Section \ref{sub:obs_jwst_spec}, \citet{dhawan_spectroscopic_2024} performed an exhaustive analysis of the SN Encore spectra and found based on the SALT3 model that at the time of the first imaging epoch ($60265$; Table \ref{tab:obs}), image 1a of the SN was $\sim28.7^{+2.0}_{-1.5}$ rest-frame days ($85^{+5.9}_{-4.4}$ observer-frame days) after peak brightness. Using the SN IDentification (\texttt{SNID}) tool \citet{dhawan_spectroscopic_2024} inferred $29.0\pm5.0$ rest-frame ($85.6\pm14.8$ observer-frame) days, entirely consistent with SALT3. As the phase inference from the combination of both epochs was more robust with \texttt{SNID}, we proceed with this constraint as a weak prior on phase. Our imaging observations therefore span $\sim70$-$156$ observer-frame ($\sim14$-$53$ rest-frame) days after peak brightness for image 1a, considering both the MJD range and $3\sigma$ uncertainty on phase. Including the relative time delay between images, we require a SN\,Ia model that extends to $\sim80$ rest-frame days to ensure we cover the fully plausible parameter space. In this paper we do not attempt to measure the image 1c delay, which will require a template image for more accurate photometry and a SN\,Ia model that extends beyond $\sim100$ rest-frame days, based on our mass models (S25). 

The only trained model that extends to $>80$ rest-frame days after peak brightness is the ``Hsiao'' spectral template \citep{hsiao_k_2007}, but this model does not capture the diversity of SNe\,Ia and instead represents a fiducial, ``normal'' SN\,Ia (i.e., nominal light curve shape and color). Based on the work of P24 and \citet{dhawan_spectroscopic_2024}, SN Encore is a fast declining SN ($15$-day rest-frame change from peak brightness in the $B$ band is $\Delta m_{15,B}\sim1.34$) that would not be well-represented by the fiducial Hsiao template. However, the BayeSN light curve model \citep{mandel_hierarchical_2022,ward_relative_2023,grayling_scalable_2024} uses the Hsiao model as a reference template, and was extended to $85$ days after peak brightness for \citet{pierel_jwst_2024}, where the extrapolation and training methods are described in detail. BayeSN is the only well-tested SN\,Ia light curve model that has both near-infrared and late-time coverage ($>+50$ days after peak brightness).


\begin{table*}[!t]
    \centering
    \caption{\label{tab:fitting} Summary of fits for SN Encore, with the fit used as our final result used in Section \ref{sec:results} shown in bold.}
    
    \begin{tabular*}{\linewidth}{@{\extracolsep{\stretch{1}}}*{7}{c}}
\toprule        
Photometry (Section)&Method (Section)&$\theta$&$E(B-V)_{\rm host}$&$t_{pk,1a}$&$\mu_{1b}/\mu_{1a}$&$\Delta t_{1b,1a}$\\
&&&(mag)&(days)&(days)\\
\hline
\texttt{phot\_bkg} (\S\ref{sub:phot_bkg})&\textsc{Glimpse} (\S\ref{sub:fitting_gausn})&$0.6_{-1.0}^{+0.8}$&$<0.1$&$60167.9_{-10.6}^{+9.0}$&$2.0_{-0.4}^{+0.5}$&$-37.3^{+13.1}_{-12.5}$\\
\texttt{phot\_bkg} (\S\ref{sub:phot_bkg})&\sntd (\S\ref{sub:fitting_sntd_bkg})&$1.4_{-0.8}^{+0.7}$&$<0.1$&$60217.3_{-7.2}^{+7.3}$&$2.6_{-0.5}^{+0.6}$&$-47.0^{+16.6}_{-13.2}$\\
\textbf{\texttt{phot\_starred} (\S\ref{sub:phot_starred})}&\textbf{\sntd (\S\ref{sub:fitting_sntd_starred})}&$\mathbf{1.1_{-0.2}^{+0.2}}$&$<0.1$&$\mathbf{60201.1_{-2.3}^{+2.1}}$&$\mathbf{1.5_{-0.5}^{+0.5}}$&$\mathbf{-39.8^{+2.9}_{-2.5}}$\\

\end{tabular*}
\tablecomments{Columns are: The photometry used from Section \ref{sec:phot}, the fitting method from Section \ref{sec:fitting}, the best-fit BayeSN model parameters $\theta$ and host $E(B-V)$, the best-fit time of peak for the model using image 1a as the reference, the magnification ratio between images 1a and 1b, and relative time delay between images 1a and 1b. Note that these uncertainties do not include systematics from Section \ref{sec:sys}.}
\end{table*}

\subsection{Time Delays with Glimpse}
\label{sub:fitting_gausn}
We first estimate the time delay of image 1b relative to image 1a using the GausSN Light curve Inference of Magnifications and Phase Shifts, Extended (\textsc{Glimpse}) model (E. Hayes et al., in preparation), which builds upon the \textsc{GausSN} model presented in \citet{Hayes_2024}. \textsc{Glimpse} estimates the time delay of lensed SNe using SN light curve templates to capture the true underlying shape of the light curve with a novel flexible, non-parametric Gaussian Process \citep[GP; see e.g.,][]{Rasmussen_2006} model for the effects of chromatic microlensing. Simultaneously with the light curve template and chromatic microlensing parameters, the model also accounts for dust extinction in the host galaxy and in the lens galaxy, for which the extinction is allowed to vary for each lensed image, based on dust laws as implemented in \texttt{sncosmo} \citep{Barbary_2022}. A more detailed description of the fitting method will be presented in E. Hayes et al. (in preparation). As \textsc{Glimpse} does not have the capability to fit for the additional systematic offsets present in the \texttt{phot\_starred} data (see Section \ref{sub:phot_starred} and \ref{sub:fitting_sntd}), we fit only the \texttt{phot\_bkg} data with this method, in order to guard against potential bias by using only a single fitting method (i.e., SNTD).

The underlying SED for the \textsc{Glimpse} fit is the above described \textsc{BayeSN} 85 day model as implemented in \texttt{sncosmo}. This implementation of the model has three free parameters to sample over: the time of maximum flux, $t_{0}$, the light curve shape parameter, $\theta$, and an ``amplitude'' parameter, which is related to the distance modulus. Compared to the nominal \textsc{BayeSN} model, this implementation does not include any residual time- and wavelength-independent variations, $\bm{\epsilon}(t,\lambda)$. Given the limited data from SN Encore, we would not expect to constrain these parameters. We assume a \citet{Fitzpatrick_1999} dust law with $R_{V} = 2.6$ for both host and lens dust effects. Given the sparsity of data for SN Encore, we do not expect the dust law shape parameter to be well-constrained, so we fix $R_{V} = 2.6$ as similar values have been reported as the population mean in a number of samples of (mainly low-$z$) SNe~Ia, including works using \textsc{BayeSN} \citep[e.g.,][]{thorp_testing_2021, ward_relative_2023}. We note that \citet{pierel_jwst_2024} found little dependence of the time delay on the choice of $R_{V}$. For SN Encore, we sample the posterior over 11 hyperparameters of the system: 3 GP microlensing hyperparameters, 3 light curve template parameters, 3 dust extinction parameters, 1 relative time delay, and 1 relative magnification.

As in \citet{Hayes_2024}, we sample the posterior using a nested sampling algorithm \citep{Neal_2003, Skilling_2004, Skilling_2006, Handley_2015a, Handley_2015b} as implemented in \texttt{dynesty} \citep{Speagle_2020}. In addition to the Gaussian prior on $t_{0}$ from \citet{Dhawan_2024}, a uniform prior on the time delay of image A relative to image B of $-$250 to 250 days, and a uniform prior on the magnification of image A relative to image B of 0.01 to 10. Varying the upper and lower bounds on the priors over the relative time delay and magnification within reason has no effect on the resulting posterior. 

We find a time delay of $\Delta t_{\rm 1b,1a} = -37.3_{-12.5}^{+13.1}$ days. The relative magnification is found to be $\beta_{\rm 1b,1a} = 2.0_{-0.4}^{+0.5}$, with image 1b being the brighter image. We do not find any evidence for microlensing affecting either image, though limited data do not enable a tight constraint on the shape of microlensing effects across time and wavelength. We constrain E(B-V)$_{\textrm{host}} < 0.07 \, (0.12)$, E(B-V)$_{\textrm{lens, im 1a}}<0.23 \, (0.40)$ and E(B-V)$_{\textrm{lens, im 1b}}<0.27 \, (0.48)$ at the 68\% (95\%) confidence level. The dust extinction in the host is found to be minimal and in the lens appears to be consistent along both lines-of-sight. 




\begin{figure}
    \centering
    \includegraphics[width=0.99\linewidth]{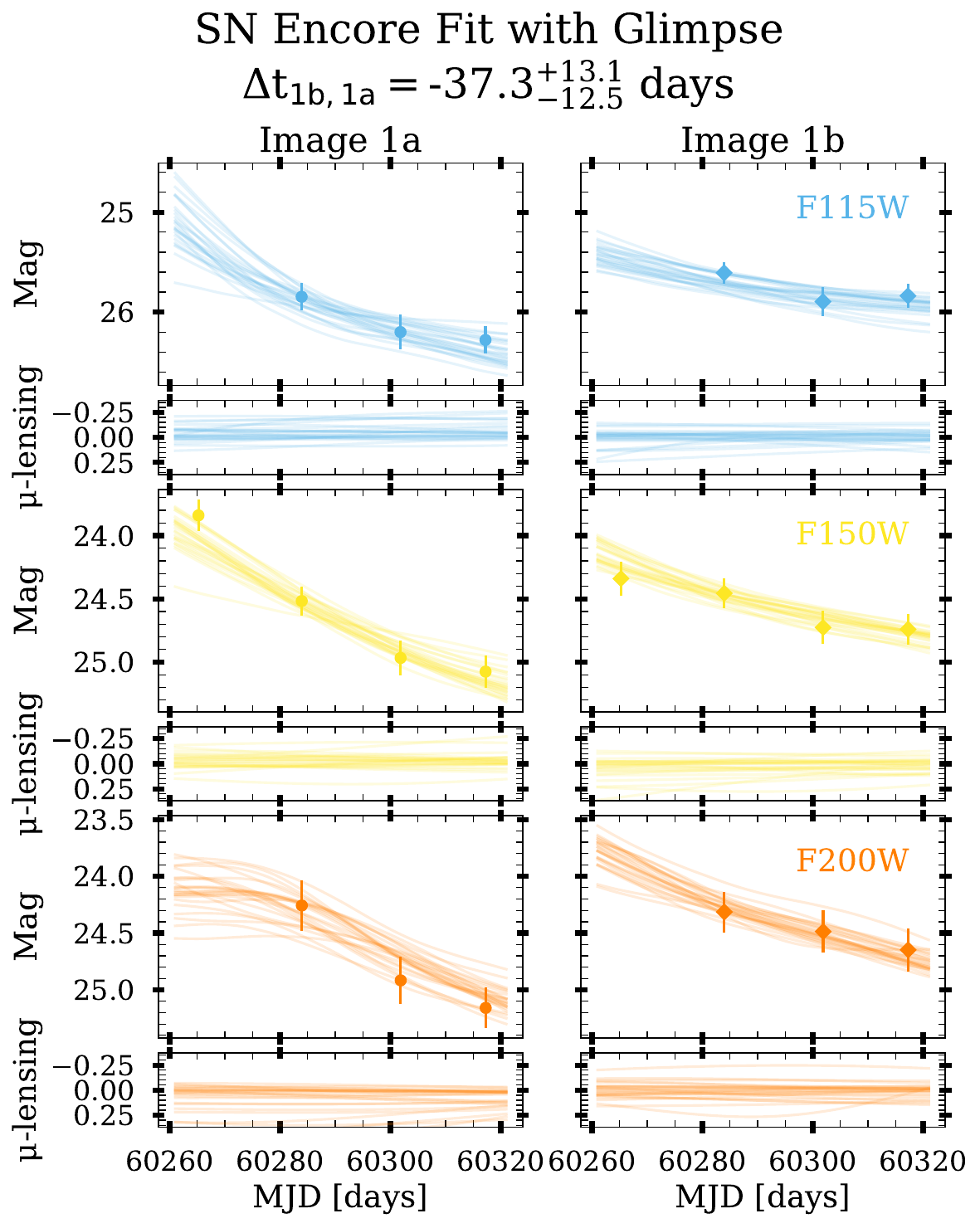}
    \caption{The \textsc{Glimpse} fit to the SN Encore image 1a and 1b light curves. Each line represents a sample from the posterior, with a total of 25 samples shown for clarity. The upper panels for each filter shows the combined template+chromatic microlensing fit to the data, while the lower panel shows only the microlensing effect in each filter.}
    \label{fig:gaussn2-fit}
\end{figure}

\subsection{Time Delays with \sntd}
\label{sub:fitting_sntd}
We also measure the time delay using the SN Time Delays (\sntd) software package \citep{pierel_turning_2019}. Previously, the \sntd ``Color'' fitting method has been used, which measures time delays directly using color curves \citep[e.g.,][]{pierel_jwst_2024}. This removes the effects of achromatic microlensing, macro magnification, and millilensing from the time-delay measurement. However, we find that in the observed phase range for \lensedsn, the expected SN\,Ia color curves are $\sim$linear in the observed filters, making a robust time-delay measurement extremely difficult. We therefore proceed with the \sntd ``Series'' method, which attempts to reconstruct the combined \lensedsn light curve using a single intrinsic model, while varying the relative time delay and magnification ratio between images. As is apparent in the corner plots shown in Figure \ref{fig:sntd_corner} and in the appendix, this leads to a strong correlation with magnification ratio that can be broken in the future with color curve fits to improved photometry. We do check that measuring time delays with the \sntd Color method is consistent with the Series method, and find that the results are nearly the same (well within $1\sigma$), but with very large uncertainties in the Color method. We attempt to account for the additional impact of (chromatic) microlensing and millilensing in Section \ref{sec:sys}. In this section we also apply Milky Way dust extinction ($E(B-V)=0.014$,  $R_V=3.1$) based on the maps of \citet{schlafly_measuring_2011} and extinction curve of \citet{fitzpatrick_correcting_1999}. In each case below we once again use the BayeSN light curve model, described in Section \ref{sub:fitting_models}, and account for the BayeSN model uncertainties when quoting uncertainties in this section. We also note that \textsc{Glimpse} has significant flexibility between its GP methodology and analytic microlensing fitting routine, and we therefore imposed a $t_{\rm{pk,1a}}$ prior based on the spectroscopic age constraint of $60180\pm15$ days. This leads to a significant difference in the $t_{\rm{pk,1a}}$ measurements in Table \ref{tab:fitting} (as \sntd imposes no such prior), but we note that the time delay and other light curve parameters seem fairly robust to this difference given the large uncertainties. This is likely due to the relatively flat behavior of the light curve in this phase range for the three filters used for fitting, which nevertheless leads to a consistent time-delay measurement (Figures \ref{fig:gaussn2-fit}, \ref{fig:sntd_bkg}).

\subsubsection{\sntd: Fitting \texttt{phot\_bkg} Photometry}
\label{sub:fitting_sntd_bkg}
\begin{figure}
    \centering
    \includegraphics[width=.75\linewidth]{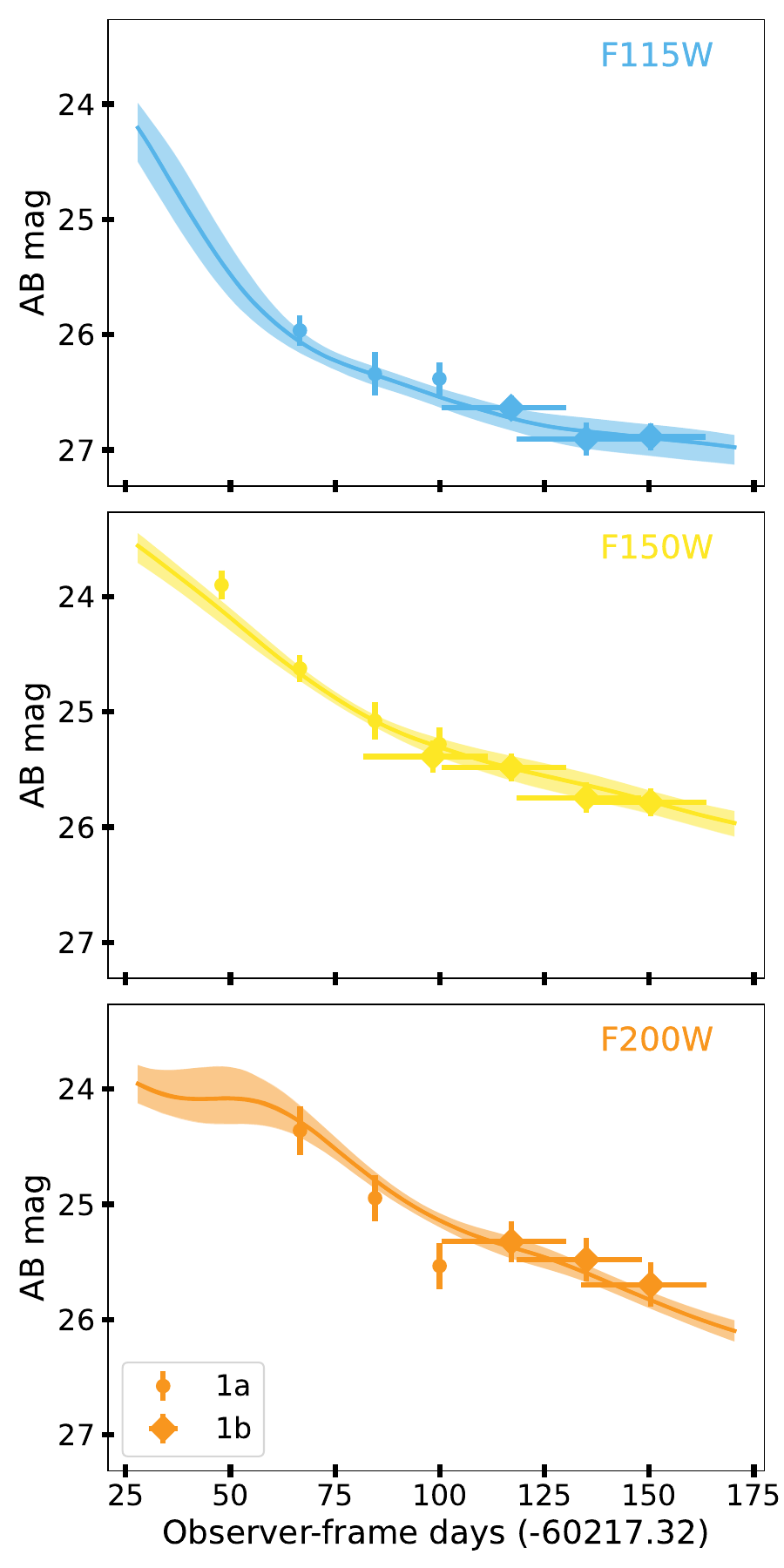}
    \caption{The best-fit \sntd result for the \texttt{phot\_bkg} dataset, described in Section \ref{sub:fitting_sntd_bkg}. Horizontal error bars represent the time-delay uncertainty.}
    \label{fig:sntd_bkg}
\end{figure}

We first fit the \texttt{phot\_bkg} photometry in order to check consistency with the \textsc{Glimpse} method above and the results of the next section. This set of photometry, shown in the appendix, only contains the \textit{JWST} short-wavelength filters (F115W, F150W, F200W), so constraints are expected to be quite uncertain, as was the case in the previous section. The results of this fitting routine are summarized in Table \ref{tab:fitting}, with the time delay in $\lesssim1\sigma$ agreement with \textsc{Glimpse}. A plot of the best-fit model to the \texttt{phot\_bkg} data is shown in Figure \ref{fig:sntd_bkg}.

\subsubsection{\sntd: Fitting \texttt{phot\_starred} Photometry}
\label{sub:fitting_sntd_starred}
\begin{figure}
    \centering
    \includegraphics[trim={0cm 0cm 0cm 0cm},clip,width=\linewidth]{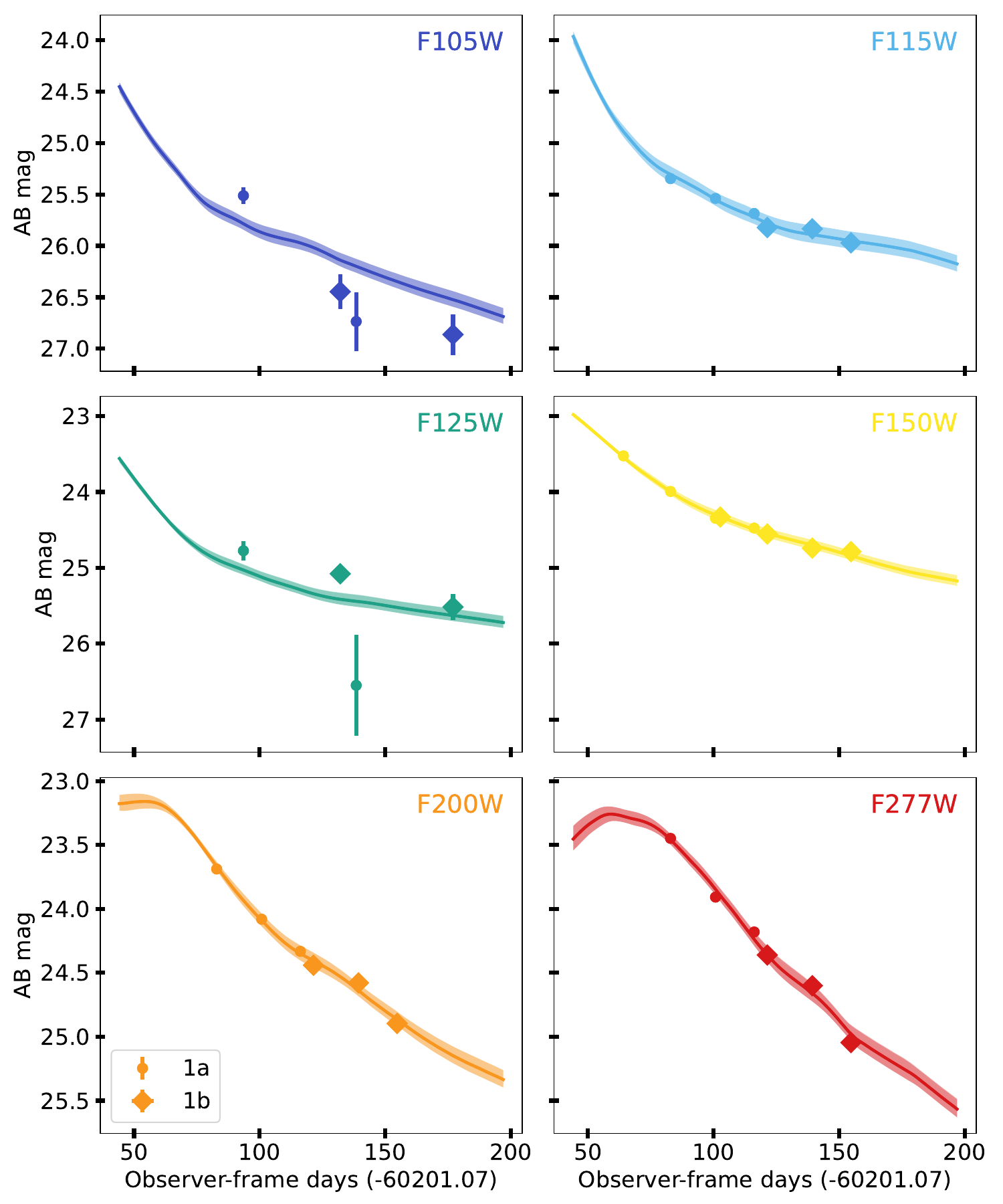}
    \caption{The best-fit \sntd result for the \texttt{phot\_starred} dataset, described in Section \ref{sub:fitting_sntd_starred}. Horizontal error bars (smaller than the points) represent the statistical-only time-delay uncertainty.}
    \label{fig:sntd_starred}
\end{figure}

The above sections give us confidence that our method is not heavily dependent on light curve model or fitting method, and now we repeat the fitting of the preceding section using the measured \texttt{phot\_starred}. The advantage of \texttt{phot\_starred} is that the method was able to extract \lensedsn fluxes robustly in all filters except for F444W. Given the anticipated phase of \lensedsn discussed in Section \ref{sub:fitting_models}, these redder filters provide critical information from the secondary infrared maximum in SNe\,Ia \citep[e.g.,][]{burns_carnegie_2018,dhawan_measuring_2018,avelino_type_2019,pierel_salt3nir_2022,mandel_hierarchical_2022}, theoretically providing a much more precise measurement of the time delay. However, as discussed in Section \ref{sub:phot_starred}, the caveat is that the lack of a template image introduces a systematic uncertainty into each filter. The uncertainty is constant in time though, which means it is possible to add these shifts (additive in flux) into the fitting routine of \sntd and constrain them alongside the other parameters in Table \ref{tab:fitting}. We did test this concept by fitting for the offsets in a simple set of simulations and found that the method was successful, but it is fairly dependent on the accuracy of the underlying SED model, BayeSN. The full corner plot for the fit is given in the appendix, showing that these offsets are robustly retrieved by the nested sampling routine. However, these extra parameters still increase the uncertainty in the time-delay measurement, meaning that a repeat analysis with a template image should still substantially reduce the final uncertainty for $H_0$. The results from this method feed into our $H_0$ measurement in Section \ref{sec:results}, and are summarized in Table \ref{tab:fitting}. The posterior distribution for the most relevant parameters is shown in Figure \ref{fig:sntd_corner}, with the full distribution shown in the appendix, while the best-fit model is shown in Figure \ref{fig:sntd_starred}.

\begin{figure}
    \centering
    \includegraphics[width=\linewidth]{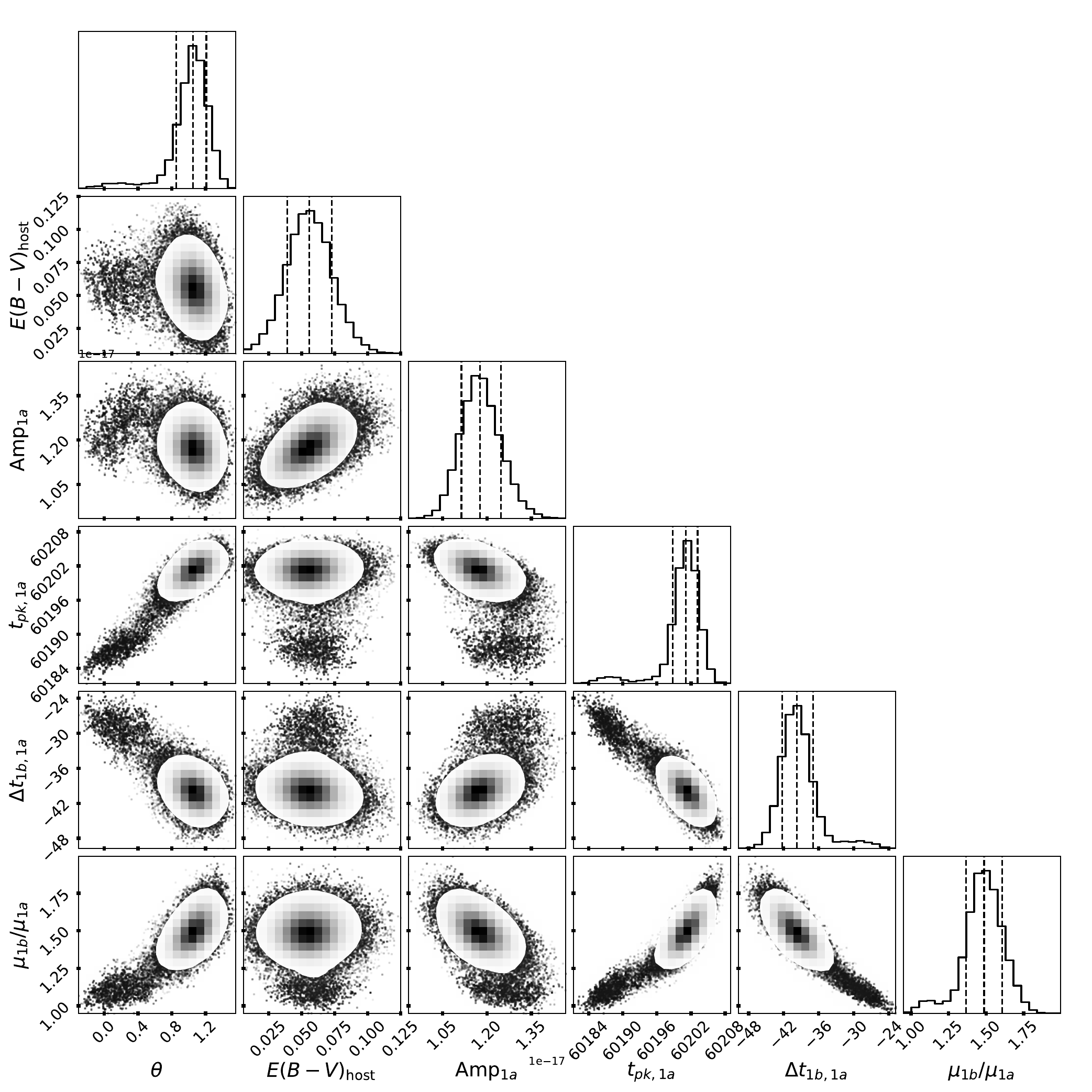}
    \caption{The posterior distribution for the most relevant parameters from the \sntd result for the \texttt{phot\_starred} dataset. The full posterior including systematic offsets is shown in the appendix.}
    \label{fig:sntd_corner}
\end{figure}

\subsubsection{Magnifications}
\label{sub:sntd_fitting_magnifications}
We follow the method outlined in \citet{pierel_jwst_2024} to measure absolute magnifications for \lensedsn. The process is to take the sample of high-$z$ SN\,Ia distance measurements and fit their distance moduli using a non-cosmological method (e.g., a simple polynomial), and then use that fit to obtain an estimate of a normal, non-lensed SN\,Ia distance modulus (DM) at $z=1.95$. By comparing this to the measured DM for \lensedsn, we obtain a measurement of the enhanced brightness due to the strong lensing magnification. Since the method was used for SN\,H0pe, additional distance measurements have been made with $z\sim2$ SNe\,Ia have improved the interpolation method in the region of the \lensedsn redshift, and we find that the polynomial method described there best represents the updated data. Proceeding with this, we find that a non-lensed SN\,Ia DM at $z=1.95$ should be $\rm{DM}=45.7$. The BayeSN model directly yields a DM from the fitting routine, which for \lensedsn is $\rm{DM}_{1a}= 42.3_{-0.1}^{+0.1}$, $\rm{DM}_{1b}=41.9_{-0.1}^{+0.1}$, which corresponds to magnifications of $\mu_{1a}=21.8_{-1.3}^{+1.2}$, $\mu_{1b}=32.4_{-2.2}^{+2.4}$.

\section{Additional Systematic Uncertainties}
\label{sec:sys}
\begin{table}[h!]
    \centering
    \caption{\label{tab:systematics} The impact of the additional explored systematics on measured parameters.}
    
    \begin{tabular*}{\linewidth}{@{\extracolsep{\stretch{1}}}*{4}{c}}
\toprule
Systematic (Section)&$\delta(\Delta t_{1b,1a})$&$\delta(\mu_{1a})$&$\delta(\mu_{1b})$\\
&(days)&&\\
\hline
Microlensing (\S\ref{sub:sys_micro}) & $_{-1.9}^{+1.3}$&$_{-4.2}^{+9.3}$ &$_{-5.9}^{+7.1}$ \\
Millilensing (\S\ref{sub:sys_milli}) & -- & $_{-6.2}^{+6.2}$&$_{-9.2}^{+9.2}$ \\ 
\hline
\hline
    \end{tabular*}
\end{table}

\subsection{Chromatic Microlensing}
\label{sub:sys_micro}
Microlensing from small perturbers such as stars in the lens plane is well known to potentially impact time-delay measurements \citep[e.g.,][]{dobler_microlensing_2006,goldstein_precise_2018,pierel_turning_2019,huber_strongly_2019,huber_holismokes_2022}. The specific intensity profiles for SNe vary in different filters, therefore as the expanding SN shell interacts with the microlensing caustics, the resulting magnification is (often) chromatic \citep{goldstein_precise_2018,foxley-marrable_impact_2018,huber_strongly_2019}. For SNe\,Ia, microlensing is essentially achromatic in the first $\sim$3 rest-frame weeks after explosion \citep[i.e., the ``achromatic phase'',][]{goldstein_precise_2018,huber_holismokes_2021}, but our observations are not within this time-frame, meaning they are plausibly impacted by chromatic (and/or achromatic) microlensing.


\begin{table}[h!]
    \centering
    \caption{\label{tab:micro_params}Parameters used to create microlensing magnification maps for each SN image, from the \citet{ertl_cosmology_2025} \texttt{iso\_halo} model.}
    
    \begin{tabular*}{\linewidth}{@{\extracolsep{\stretch{1}}}*{4}{c}}
\toprule
Image&$\kappa$&$\gamma$&$s$\\
\hline
1a& $0.75$&$0.31$&$0.95$\\
1b&$0.66$&$0.30$&$0.7$\\
\hline
\hline
    \end{tabular*}

\end{table}
We produced a magnification map for each SN image and convolved the SN\,Ia light profiles from four theoretical models \citep{suyu_holismokes_2020,huber_holismokes_2021,pierel_jwst_2024} with each magnification map ($1000$ random positions per model for a total of $4\,000$ convolutions per map)  in the manner of \citet{huber_strongly_2019}. The parameters used to generate the magnification maps are the lens surface-mass density scaled to the critical value ($\kappa$), the image shear ($\gamma$), and the smooth matter fraction ($s$). Here $\kappa$ and $\gamma$ at each SN image position are those of \citet{ertl_cosmology_2025} (little variation is seen in the S25 compilation of lens models), and we assume a value of 
$s=0.95$ for image 1a, and $s=0.7$ for image 1b, given the presence of a foreground galaxy located $\sim2\arcsec$ from image 1b. 
Table \ref{tab:micro_params} shows the $\kappa$ and $\gamma$ values used for each SN image. We selected $1000$ random chromatic microlensing curves and varied the flux predicted by a baseline BayeSN model (Section \ref{sub:fitting_models}) in each band by the phase- and wavelength-dependent magnification curves. Finally, we ran the same \sntd method used in Section \ref{sec:fitting} on each of the realizations, to produce a distribution of recovered time delays, from which we remove outliers (due to extreme microlensing events) using a simple $\sigma$-clip. We find that the microlensing plausibly impacting image 1b is likely more extreme than image 1a, though the phase range of image 1a seems to yield more chromatic effects. The summary of results is given in Table \ref{tab:systematics}, but we note that the vast majority of the uncertainty in $\Delta t_{1b,1a}$ comes from the uncertainty in the reference time of peak brightness for image 1a, meaning that a future $\Delta t_{1b,1c}$ (or $\Delta t_{1b,1d}$) delay would be free of much of this uncertainty. We include this systematic distribution as a systematic uncertainty in our time delay distribution. 

\subsection{Millilensing}
\label{sub:sys_milli}

In addition to the effects of microlensing, we may also consider perturbations to the smooth lens model magnifications due to the presence of dark matter sub-halos. We follow the procedure outlined by \citet{larison_lenswatch_2025}, using the open-source Python package, \textsc{pyHalo}\footnote{\url{https://github.com/dangilman/pyHalo}} \citep{gilman_warm_2020}, to simulate dark matter substructures around and along the lines of sight (LOSs) to each \lensedsn image (1a, 1b). We use the smooth lens model results from the \texttt{iso\_halo} mass model (Table \ref{tab:micro_params}) and image positions presented by \citet{ertl_cosmology_2025} to run $1000$ realizations of a simulation that places dark matter sub-halos uniformly out to a radius of $1\arcsec$ around each SN image in the lensing plane. Additionally, dark matter sub-halos are generated in a cylinder along the LOSs to the images and projected back to the source position, with a constant physical radius set by the comoving distance to the lens multiplied by $1\arcsec$. The lens-plane and LOS sub-halos are generated following a power-law and Sheth-Tormen \citep{sheth_large-scale_1999} mass function respectively, with masses ranging from $10^{5.5} - 10^9 M_{\odot}$ and Navarro-Frenk-White (NFW) mass profiles \citep{navarro_universal_1997}. To set the overall number of sub-halos generated, we assume a value for the convergence due to the presence of sub-halos that is equal to some percent of the smooth lens model value, which we call $f_{\rm sub}$. To test a range of possibilities that are supported by both theory and observation \citep{dalal_tests_2004,gilman_warm_2020}, we run three sets of simulations with $f_{\rm sub} = 1\%,\, 5\%$, and $10\%$.

After outlier removal, we find that the additional uncertainty for the median case ($f_{\rm sub} = 5\%$), is around $29\%$ for images 1a \& 1b. This uncertainty does not account, however, for the large extended tail toward larger magnifications, which indicates it is possible that one or more massive sub-halos could substantially perturb the observed image magnifications. These uncertainties are much larger than for previous strongly lensed supernovae, with \citet{kelly_constraints_2023} finding that millilensing added an additional $10\%$ uncertainty in the magnifications of SN~Refsdal, while \citet{pierel_jwst_2024} found an additional uncertainty of $4\%$ - $8\%$ for SN~H0pe. \citet{larison_lenswatch_2025} found even less of an effect, but for a galaxy-scale system. We caution that all of these uncertainties (including for SN Encore) are likely overestimated due to ignoring the effects of tidal stripping, which may significantly reduce the subhalo population and thus the extent of millilensing \citep{Du_Tidal_2025}.

Although these additional uncertainties do not affect the time-delay measurements, as millilensing is an achromatic and time-independent effect, they do negatively impact the use of \lensedsn to break the mass-sheet degeneracy in this lensing system. With template observations, precise estimates of the the image magnifications of \lensedsn will be possible due to its standardizable nature; however, producing a smooth lens model that predicts these observed magnifications may become difficult given the additional magnification due to unconstrained millilensing and/or microlensing. Adopting a method that includes these effects when fitting for the mass model, such as done by \citet{arendse_zwicky_2025} with microlensing in the galaxy scale systems: iPTF16geu and SN Zwicky, could potentially still allow for the standardizable candle nature of \lensedsn to be leveraged in the lens modeling process.

\section{Measurement of the Hubble Constant}
\label{sec:results}

\begin{table}[h!]
    \centering
    \caption{\label{tab:final_delay_mag} Final time delays and magnifications, with systematic uncertainties, measured in Section \ref{sec:fitting} and Section \ref{sec:sys}, respectively.}
    
    \begin{tabular*}{\linewidth}{@{\extracolsep{\stretch{1}}}*{3}{c}}
\toprule
$\Delta t_{1b,1a}$&$\mu_{1a}$&$\mu_{1b}$\\
(days)&&\\
\hline
$-39.8^{+3.9}_{-3.3}$& $21.8_{-8.6}^{+9.3}$&$32.4_{-11.0}^{+11.1}$\\
\hline
\hline
    \end{tabular*}

\end{table}

We now obtain a measurement of $H_0$ by combining the final result on the time delay, $\Delta t_{1b,1a}$, with the seven mass models produced in S25. While each mass model is described in detail in S25, we quickly summarize the models here and then describe the method of weighting the $H_0$ prediction from each model for the $H_0$ inference. 

\subsection{Mass Models}
\label{sub:h0_models}
There are a total of six lens modeling methods, which produced a total of seven lens models, presented in S25 to be used in this analysis. Details of each lens model, the selection of constraints for all models, and the final weighting scheme are given in S25. The models are summarized here in Table \ref{tab:h0}, including the S25 derived weights and values for $H_0$ given our time-delay measurement.

\subsection{Blinded Analysis}
\label{sub:h0_blinding}
We performed this analysis fully blinded to the results of the mass modeling in S25, to eliminate the impact of human bias. The blinding protocols within the lens modeling teams are described in S25, but relevant to this work is that no information about the time delay or magnification was shared between the lens-modeling teams and the time-delay teams. The time-delay measurement from \textsc{Glimpse} was not blinded from the \sntd measurements, as it was used as a cross-check to ensure we could move forward comfortably with the \sntd result. We note that while such a blinded analysis can only truly happen once for a given dataset, a future updated measurement leveraging a template image can still use blinded offsets to similarly guard against potential bias. 

\subsection{Mass Model Weighting}
\label{sub:h0_weights}
\begin{figure*}[ht]
    \centering
    \includegraphics[width=0.6\linewidth]{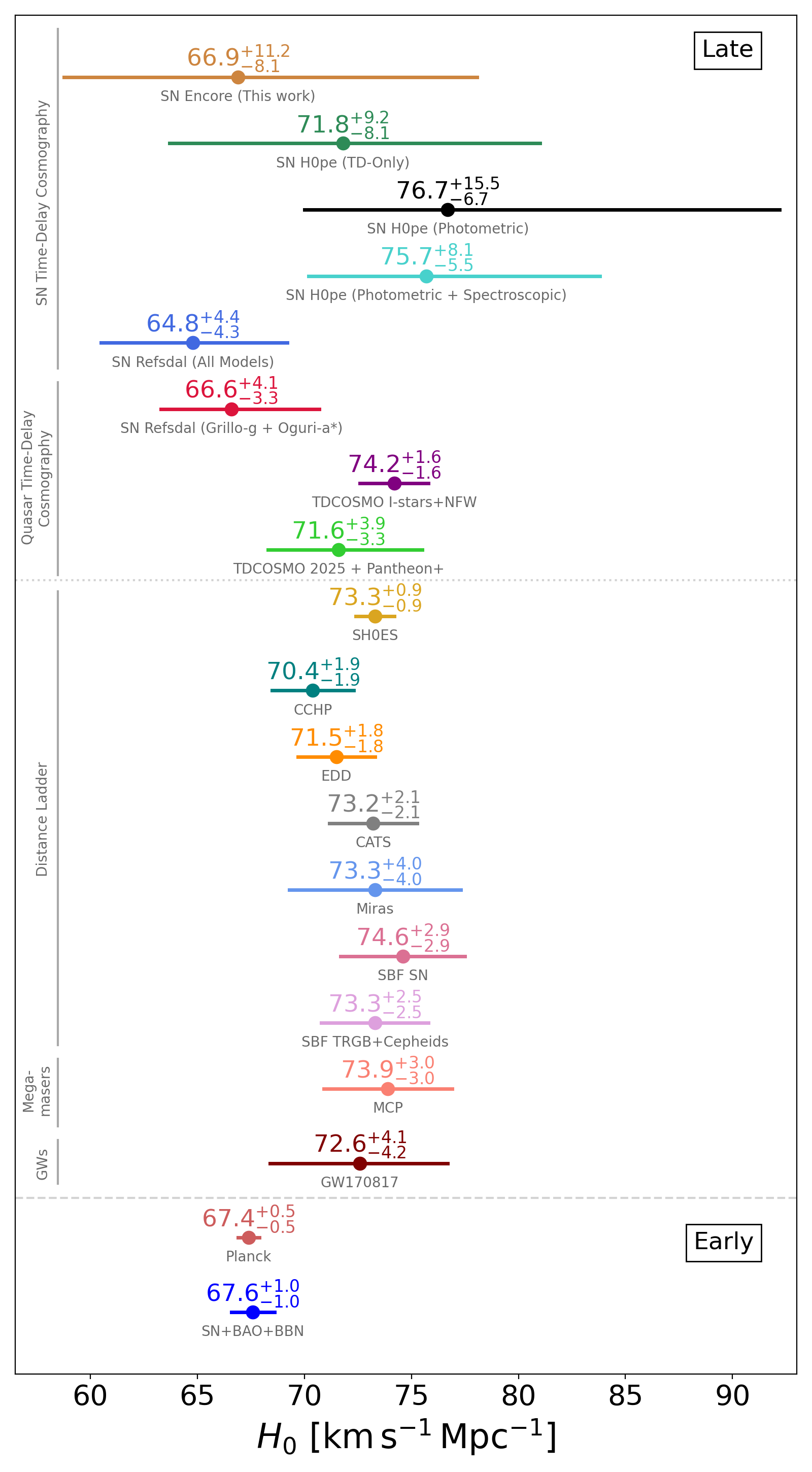}
    \caption{Measurements of $H_0$ at the time of this publication, with details and references given in Section \ref{sec:intro}. The measurement in this work is shown at the top, labeled ``SN Encore (This work)''. This figure is adapted from \citet{pascale_sn_2025} and \citet{bonvin_h0licow_2020}.}
    \label{fig:h0_current}
\end{figure*}

\begin{table*}[!t]
    \centering
    \caption{\label{tab:h0} Summary of mass model results for the $H_0$ inference using \lensedsn.}
    
    \begin{tabular*}{\linewidth}{@{\extracolsep{\stretch{1}}}*{4}{c}}
\toprule        
\#&Mass Model&Weight&$H_0$\\
&&&$(\rm{km}\,\rm{s}^{-1}\rm{Mpc}^{-1})$\\
\hline
1&glafic&$1.47\times10^{-1}$&$59.5_{-6.6}^{+9.8}$\\
2& GLEE&$8.45\times10^{-1}$&$67.7_{-7.6}^{+11.0}$\\
3& Lenstool I&$2.18\times10^{-19}$&$106.4_{-35.8}^{+  31.8}$\\
4&Lenstool II&$1.72\times10^{-4}$&$66.4_{-8.7}^{+14.6}$\\
5&MrMARTIAN&$7.69\times10^{-3}$&$76.0_{-7.8}^{+11.5}$\\
6&WSLAP+&$3.78\times10^{-463}$&$88.2_{-18.8}^{+26.5}$\\
7&Zitrin-analytic  &$2.61\times10^{-30}$&$ 77.8_{-16.7}^{+23.9}$\\
\hline
\textbf{Combined}&--&--&$\mathbf{66.9^{+11.2}_{- 8.1}}$\\

\end{tabular*}
\tablecomments{Columns are (for each mass model): The mass model ID, name/reference, the maximum likelihood weight, and $H_0$ value.}
\end{table*}
Previous measurements of $H_0$ using lensed SNe have weighted the models based on their ability to reproduce observables of the lensed SN, including magnification ratios \citep{kelly_constraints_2023}, absolute magnifications \citep{pascale_sn_2025}, and the ratio of two time-delay measurements for one system for which the $H_0$ dependence cancels out \citep{pascale_sn_2025}. Here, until a template observation of the system enables robust photometry for all \lensedsn images (including 1c), we cannot leverage these methods. We have only a single time-delay measurement, and the level of uncertainty in magnification at this stage does not yield meaningful weights. We explored using the Bayesian information criterion (BIC) to weight the different mass models, since BIC accounts for both goodness-of-fit and model complexity, but found the method infeasible given the use of non-parametric models and different priors for each model. BIC weighting is still shown in S25 for comparison, but in order to fairly treat all lens models we use the maximum likelihood of the multiple lensed image positions directly for our $H_0$ constraint. Note that this is possible only because all models ingest exactly the same data. Therefore, we use the gold baseline image positions ($23$ multiple image positions with circularized positional uncertainties) described in S25 to obtain a set of mass models with directly comparable likelihoods:

\begin{equation}
    L_{\rm{impos}} = A_{\rm{impos}} \times \exp(-\chi^2_{\rm{impos}}/2),
\end{equation} 
where $A_{\rm{impos}}$ is a normalization constant and $\chi^2_{\rm{impos}}$ is the measured $\chi^2$ between a given mass model's image position predictions and the measured values.  With these definitions, the BIC becomes:
\begin{equation}
    {\rm BIC} = -2 \ln(L^*_{\rm{impos}}) + N_{\rm P} \times \ln (2N_{\rm{im}}), 
\end{equation}
where $L_{\rm impos}^*$ is the maximum likelihood of $L_{\rm impos}$, $N_{\rm P}$ is the number of freely-varying parameters in the model, and $N_{\rm{im}}=23$ is the number of multiple images used as constraints. In order to obtain $L_{\rm{impos}}^*$ for each mass model, each modeling team provided $N_{\rm S}$ samples of their lens mass models for a fixed cosmology. Then for each sample $\{1,...,N_S\}$, we draw $M$ random values of $H_0$, rescale the predicted time delays and compute the time-delay likelihood $L_{\rm \Delta t}^i$ as the weight for each draw.  
We thereby obtain $N_S\times M$ weighted samples of $H_0$ value for each model model, with the time-delay likelihood as weight.  Each of the model is then further weighted by $L_{\rm impos}^*$.
We report the weights $L_{\rm impos}^*$ (and the corresponding $H_0$) in Table \ref{tab:h0}. The final result is shown in Table \ref{tab:h0}, with a measurement of $H_0=66.9^{+11.2}_{-8.1}\,\rm{km}\,\rm{s}^{-1}\rm{Mpc}^{-1}$. This value is shown in the context of other $H_0$ measurements in Figure \ref{fig:h0_current}. 

\section{Discussion}
\label{sec:conclusion}
Here we present a joint measurement of $H_0$ with S25, using the multiply-imaged \lensedsn, which is lensed by the MACS J0138.0$-$2155 cluster and hosted by the galaxy MRG-M0138 at $z=1.95$. We measure photometry using two methods and employ two time delay fitting codes as robustness checks, with a final measured time delay of $\Delta t_{1b,1a}=$\sntdstardtba\sntdstardtbaerr\, days. S25 provides the mass modeling necessary to infer $H_0$ when combined with our time-delay measurement, with a total of seven different mass models. The mass modeling and time-delay measurement were made under blinded conditions, ensuring no bias from human knowledge. Using a maximum likelihood-weighting, the final result is $H_0=66.9_{-8.1}^{+11.2}\,\rm{km}\,\rm{s}^{-1}\rm{Mpc}^{-1}$. We also measure absolute magnifications for \lensedsn by comparing the observed DM to the distribution of normal non-lensed SNe\,Ia at $z=1.95$, and obtain $\mu_{1a}=21.8^{+9.3}_{-8.6}$, $\mu_{1b}=32.4^{+11.1}_{-11.0}$. These measurement uncertainties are dominated by systematics introduced by our microlensing and millilensing simulations, but nevertheless the mean values are in quite good agreement with the lens model predictions in S25 ($|\mu_{1a}|=26.9^{+2.5}_{-2.6}$, $|\mu_{1b}|=40.7^{+7.0}_{-6.9}$), possibly suggesting that the possible variability explored in each scenario could be overestimated. 

The uncertainty on $H_0$ is dominated by the time-delay uncertainty, for which the major contributors are the photometry and microlensing. The agreement between the lens model predictions and measured absolute magnifications may suggest the microlensing simulations are too conservative, an avenue that can be explored in future work. The photometric uncertainties are large primarily due to the lack of template imaging for the \textit{JWST} observations. Template observations are expected to be obtained for \lensedsn, which would drastically improve the photometric uncertainties and remove the additional fitting uncertainties associated with marginalizing over the STARRED systematics associated with a lack of template (Section \ref{sub:phot_starred}). Overall, we expect future work to reduce the uncertainty on $H_0$ by roughly a factor of $2$, but this work shows the system offers an excellent opportunity for a precision $H_0$ measurement. 

Finally, we note that the reappearance of SN\,Requiem is also expected within the next couple of years (see S25). Such a reappearance will enable an extremely high-precision estimate of the time delay given the extraordinarily long baseline, resulting in an $H_0$ inference of $\sim2$-$3\%$ based on current lens model uncertainties. The reappearance of SN\,Requiem, combined with a reanalysis of \lensedsn once template imaging becomes available, will also provide the first opportunity for a joint $H_0$ measurement of two lensed SNe from the same system. Overall, particularly considering the success of \lensedsn, the remarkable MACS J0138.0$-$2155 cluster will continue to provide unique tests for lens models and cosmology in the coming years.

\clearpage

\begin{center}
    \textbf{Acknowledgments}
\end{center}

This paper is based in part on observations with the NASA/ESA Hubble Space Telescope and James Webb Space Telescope obtained from the Mikulski Archive for Space Telescopes at STScI. We thank the DDT and JWST/HST scheduling teams at STScI for extraordinary effort in getting the DDT observations used here scheduled quickly. The specific observations analyzed can be accessed via \dataset[DOI: 10.17909/snj9-an10]{https://doi.org/10.17909/snj9-an10}"; support was provided to JDRP and ME through program HST-GO-16264. JDRP is supported by NASA through a Einstein Fellowship grant No. HF2-51541.001 awarded by the Space Telescope Science Institute (STScI), which is operated by the Association of Universities for Research in Astronomy, Inc., for NASA, under contract NAS5-26555. Support for programs JWST GO-2345 and DD-6549 was provided by NASA through a grant from STScI. GG acknowledges support from the Italian Ministry of University and Research through Grant PRIN-MIUR 2020SKSTHZ. This work was supported by JSPS KAKENHI Grant Numbers JP25H00662, JP22K21349. SC acknowledges this research was supported by Basic Science Research Program through the NRF funded by the Ministry of Education (No. RS-2024-00413036).
MJJ acknowledges support for the current research from the National Research Foundation (NRF)
of Korea under the programs 2022R1A2C1003130 and RS-2023-00219959.
MM acknowledges support by the SNSF (Swiss National Science Foundation) through return CH grant P5R5PT\_225598. AZ acknowledges support by Grant No. 2020750 from the United States-Israel Binational Science Foundation (BSF) and Grant No. 2109066 from the United States National Science Foundation (NSF); and by the Israel Science Foundation Grant No. 864/23. FP acknowledges support from the Spanish Ministerio de Ciencia, Innovación y Universidades (MICINN) under grant numbers PID2022-141915NB-C21. EEH is supported by a Gates Cambridge Scholarship (\#OPP1144). SS has received funding from the European Union’s Horizon 2022 research and innovation programme under the Marie Skłodowska-Curie grant agreement No 101105167 — FASTIDIoUS. EM and SHS thank the Max Planck Society for support through the Max Planck Fellowship for SHS. This project has received funding from the European Research Council (ERC) under the European Union's Horizon 2020 research and innovation programme (LENSNOVA: grant agreement No 771776). This work is supported in part by the Deutsche Forschungsgemeinschaft (DFG, German Research Foundation) under Germany's Excellence Strategy -- EXC-2094 -- 390783311. DG acknowledges support from a Brinson Prize Fellowship Grant. S.T.\ was supported by funding from the European Research Council (ERC) under the European Union's Horizon 2020 research and innovation programmes (grant agreement no. 101018897 CosmicExplorer), and from the research project grant `Understanding the Dynamic Universe' funded by the Knut and Alice Wallenberg Foundation under Dnr KAW 2018.0067. AA acknowledges financial support through the Beatriz Galindo programme and the project PID2022-138896NB-C51 (MCIU/AEI/MINECO/FEDER, UE), Ministerio de Ciencia, Investigación y Universidades. C.G. PR, PB acknowledge financial support through the grant PRIN-MIUR 2020SKSTHZ. J.M.D. acknowledges the support of project PID2022-
138896NB-C51 (MCIU/AEI/MINECO/FEDER, UE) Ministerio de Ciencia, Investigación y Universidade. M. J. Jee acknowledges support for the current research from the National Research Foundation (NRF) of Korea under the programs 2022R1A2C1003130 and RS-2023-00219959. J.M.D. acknowledges the support of project PID2022-
138896NB-C51 (MCIU/AEI/MINECO/FEDER, UE) Ministerio de Ciencia, In-
vestigación y Universidade. SD acknowledges support from  UK Research and Innovation (UKRI) under the UK government’s Horizon Europe funding Guarantee EP/Z000475/1.

\clearpage

\bibliographystyle{aasjournal}

\appendix
\begin{appendices}

\section{Photometry}

\begin{figure*}[h!]
    \centering
    \includegraphics[width = \textwidth]{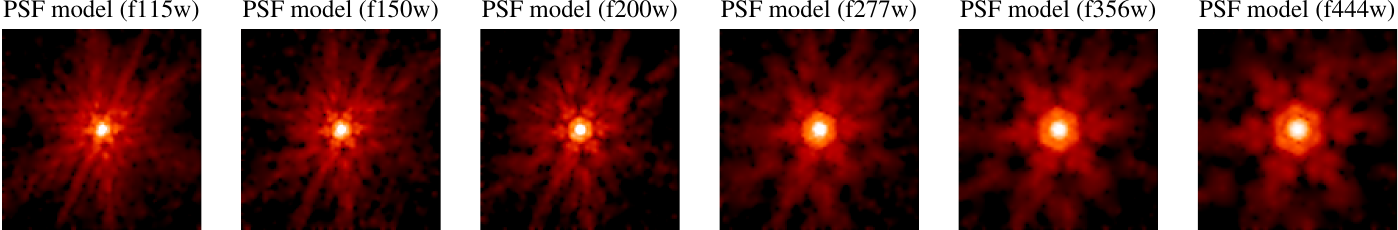}
    \caption{PSF model used for the light curve extraction. The model is fitted on 3 or 4 stars, depending on the photometric bands, using STARRED \citep{Michalewicz2023, Millon_2024}. The PSF is reconstructed on a pixel grid twice smaller than the original one. \label{fig:PSF_starred}}
\end{figure*}
\section{Time-delay Measurements}
\begin{figure*}[h!]
    \centering
    \includegraphics[width=0.8\linewidth]{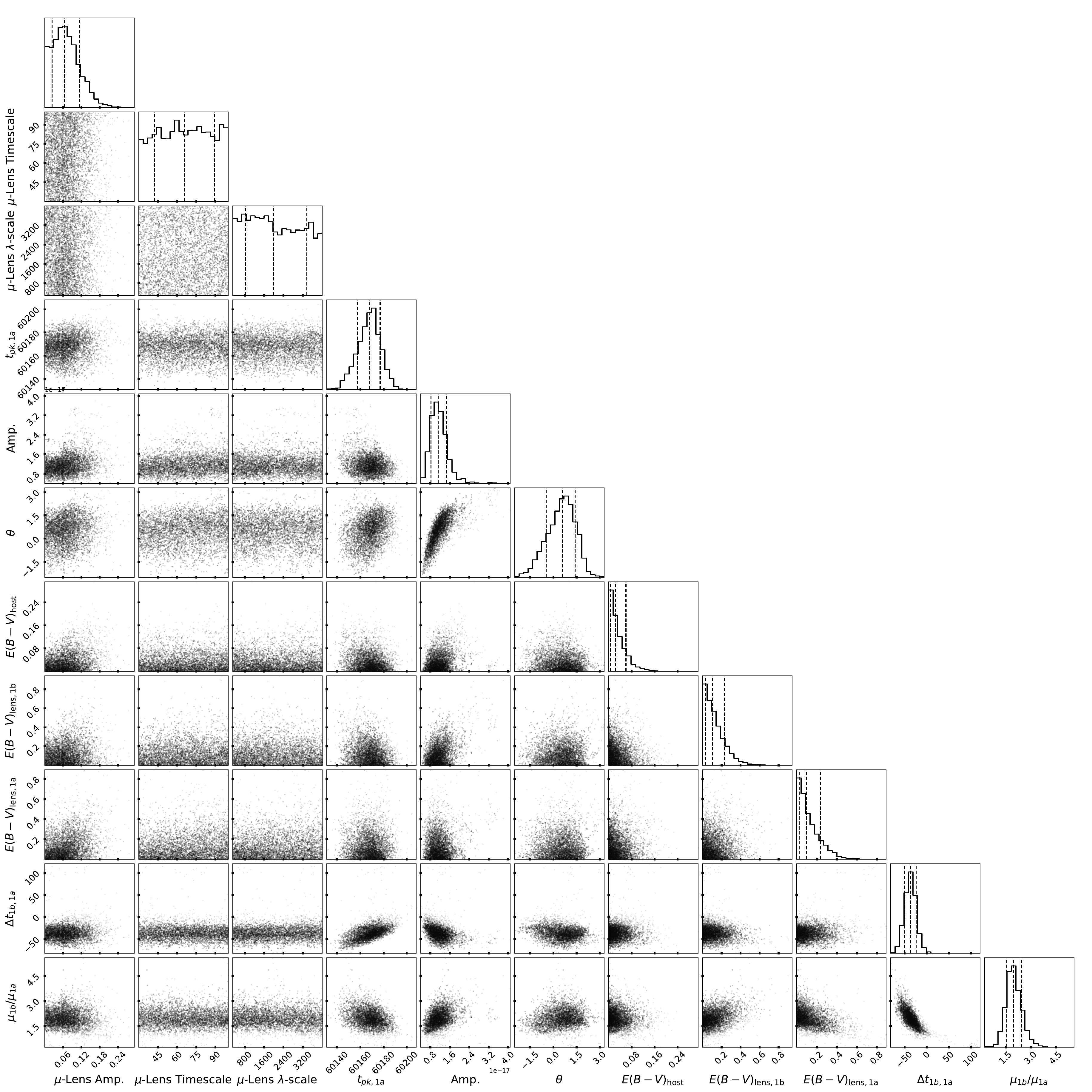}
    \caption{The posterior for the \textsc{Glimpse} fit to the SN Encore image 1a and 1b light curves.}
    \label{fig:sntd-corner-full}
\end{figure*}
\begin{figure*}[h!]
    \centering
    \includegraphics[width=0.8\linewidth]{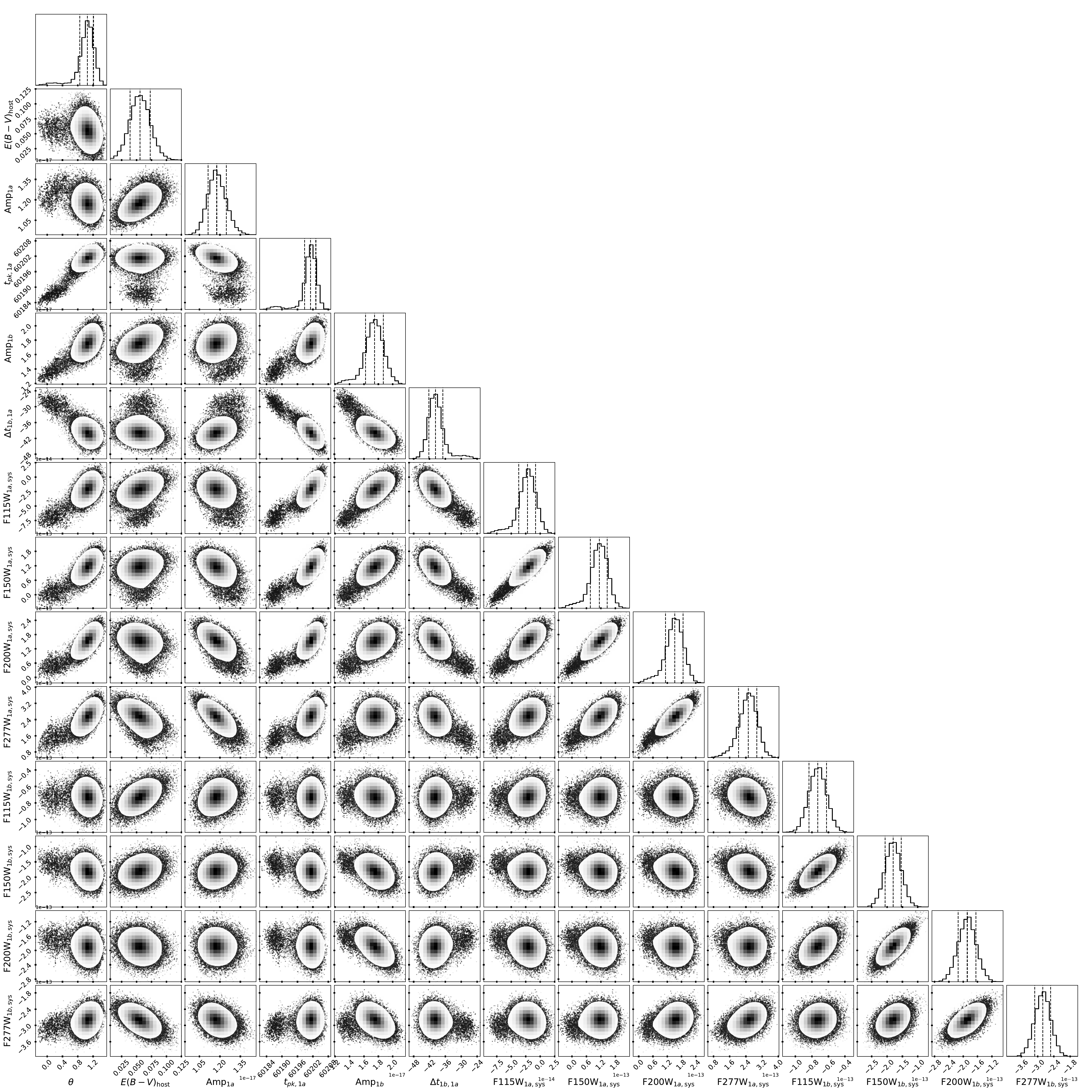}
    \caption{The posterior for the \sntd fit to the SN Encore image 1a and 1b light curves.}
    \label{fig:sntd-corner-full}
\end{figure*}
\end{appendices}
\end{document}